\documentclass[twocolumn,1p]{elsarticle}

\usepackage{lineno,hyperref}
\modulolinenumbers[5]

\usepackage{graphicx,epstopdf}
\usepackage{epsfig}
\usepackage{subfigure}
\usepackage{amsmath,amssymb,amstext}
\usepackage[export]{adjustbox}
\usepackage{float}
\newcommand{\Ei}{\mathrm{Ei}}

\newcommand{\muBB}{\mu_\text{\tiny{BB}}}
\newcommand{\muUU}{\mu_\text{\tiny{UU}}}
\newcommand{\muUB}{\mu_\text{\tiny{UB}}}
\newcommand{\muBU}{\mu_\text{\tiny{BU}}}

\newcommand{\muXY}{\mu_\text{\tiny{XY}}}
\newcommand{\muB}{\mu_\text{\tiny{B}}}
\newcommand{\muU}{\mu_\text{\tiny{U}}}

\newcommand{\deltaB}{\delta_\text{\tiny{B}}}
\newcommand{\deltaU}{\delta_\text{\tiny{U}}}
\newcommand{\pBB}{p_\text{\tiny{BB}}}
\newcommand{\pUU}{p_\text{\tiny{UU}}}
\newcommand{\pUB}{p_\text{\tiny{UB}}}
\newcommand{\pBU}{p_\text{\tiny{BU}}}
\newcommand{\LBB}{L_\text{\tiny{BB}}}
\newcommand{\LUU}{L_\text{\tiny{UU}}}
\newcommand{\LUB}{L_\text{\tiny{UB}}}

\newcommand{\NB}{N_\text{\tiny{B}}}
\newcommand{\NU}{N_\text{\tiny{U}}}

\begin{document}
\begin{frontmatter}

\title{Biased-voter model: how persuasive a small group can be? }

\author{Agnieszka Czaplicka$^{(1)}$, Christos Charalambous$^{(2)}$, Raul Toral$^{(2)}$, Maxi San Miguel$^{(2)}$}
\address{(1) Center for Humans and Machines, Max Planck Institute for Human Development, Lentzeallee 94, Berlin 14195, Germany.\\
(2) Institute for Cross-Disciplinary Physics and Complex Systems IFISC (CSIC-UIB), Campus UIB, 07122 Palma de Mallorca, Spain}

\date{\today}

\begin{abstract}
We study the voter model dynamics in the presence of confidence and bias. We assume two types of voters. Unbiased voters whose confidence is indifferent to the state of the voter and biased voters whose confidence is biased towards a common fixed preferred state. We study the problem analytically on the complete graph using mean field theory and on an Erd\H{o}s-R\'enyi random network topology using the pair approximation, where we assume that the network of interactions topology is independent of the type of voters. We find that for the case of a random initial setup, and for sufficiently large number of voters $N$, the time to consensus increases proportionally to $\log(N)/\gamma v$, with $\gamma$ the fraction of biased voters and $v$ the parameter quantifying the bias of the voters ($v=0$ no bias). We verify our analytical results through numerical simulations. We study this model on a biased-dependent topology of the network of interactions and examine two distinct, global average-degree preserving strategies (model I and model II) to obtain such biased-dependent random topologies starting from the biased-independent random topology case as the initial setup. Keeping all other parameters constant, in model I, $\muBU$, the average number of links among biased (B) and unbiased (U) voters is varied at the expense of $\muUU$ and $\muBB$, i.e. the average number of links among only unbiased and biased voters respectively. In model II, $\muBU$ is kept constant, while $\muBB$ is varied at the expense of $\muUU$. We find that if the agents follow the strategy described by model II, they can achieve a significant reduction in the time to reach consensus as well as an increment in the probability to reach consensus to the preferred state. Hence, persuasiveness of the biased group depends on how well its members are connected among each other, compared to how well the members of the unbiased group are connected among each other.

\end{abstract}

\begin{keyword}
Biased voter model, Biased-dependent complex network topologies
\end{keyword}

\end{frontmatter}

\section{Introduction}

The process by which people adopt an opinion about a given issue, such as endorsing a political option or choosing a commercial product, is a complex social phenomenon, and often the underlying mechanisms driving opinion dynamics are not well understood. Yet, public opinion is today a key player in most issues faced by our societies and policy makers are obliged to take into account the evolution of public opinion. Broadly speaking, the potential influences on an individual's opinion can be divided in three categories: those that are intrinsic to the individual such as, for example, personal beliefs; global external factors such as mass media and, finally, interactions with other members of the society. In order to capture the opinion changes subject to this latter source of influence, and inspired by the idea of understanding macroscopic behavior emerging from simple interactions between particles, a plethora of models for opinion dynamics has emerged in the statistical physics literature~\cite{2009Castellano,2019Bouchaud,2020SMT}. Among them, one of the simplest and most extensively studied is the so-called {\slshape voter model}~\cite{2019Redner}, a model of opinion dynamics that leads to herding. It was independently proposed in various research fields to study, among other things, neutral genetic drift in an ideal population~\cite{1969Kimura,1970Crow}, competition for territory between two countries~\cite{1973Clifford}, spreading of infectious diseases~\cite{2011Pinto}, language competition~\cite{2006Castello,2010Vazquez,2010Chapel,2012Patriarca}, kinetics of catalytic reactions~\cite{1992Krapivsky,1996Frachebourg,1996BenNaim}, coarsening phenomena~\cite{1996BenNaim,2001Dornic}, opinion dynamics~\cite{2003Vazquez}, political elections~\cite{2014Fernandez}, etc. One of the attractiveness of the voter model is that it is one of the few known interacting-particle models exactly solvable in regular lattices of any spatial dimension $D$~\cite{2013Liggett,2010Liggett}. Directly, or in many variations, it can also be related to other standard well studied models in mathematics and physics such as coalescing random walkers~\cite{1975Holley}, the zero-temperature Glauber kinetic Ising model and the linear Glauber model~\cite{2005Castellano}. 

In its original formulation~\cite{1973Clifford,1975Holley}, the voter model was introduced as an Ising-like model where an individual (an ``agent'' or ``voter'') associated with a lattice site $i$ can adopt two different values or ``opinions'' $s_{i}=\pm1$. The dynamics of the system is implemented by randomly choosing one individual and assigning to it the value of the opinion of one of its randomly chosen nearest neighbors. The voter model is characterized by purely noise-driven diffusive dynamics. It exhibits two symmetric absorbing states, called ``consensus'' states, from which the system cannot escape and which, for finite-size systems, are reached almost surely. If $N$ is the total number of voters, the mean time to reach consensus, $T_N$, scales in regular lattices as $N^{2}$ in $D=1$, as $N\log N$ in $D=2$, and as $N$ in $D=3$~\cite{Krapivsky:2010}. Many social systems display interactions that find a better characterization as complex networks with distinctive connectivity properties~\cite{2001Strogatz,2002Barabasi}. For this reason, in the last decades, an extensive effort has been devoted to studying voter-like models on complex networks~\cite{2003Castellano,2005bisSuchecki,2005Suchecki,2007Castello,2008Sood,2008Vazquez,2014Iwamasa,2008Baxter}. In this scenario, defining $\mu_{k}$ as the $k$-th moment of the degree distribution, it is found that for uncorrelated networks, $T_{N}\sim N\mu_{1}^{2}/\mu_{2}$, which grows sublinearly in $N$ for a sufficiently broad degree distribution~\cite{2008Sood,2008Vazquez}. 

Although the early versions of the model consider that all agents are identical, it is obvious that in real applications there will be structural differences between the agents. For instance, the number of nearest neighbors, the rate of interactions, the preference for one or another state, etc. can broadly vary. These inhomogeneities, modeled as quenched disorder, are known to have an important relevance in non-equilibrium systems with absorbing states~\cite{1996Moreira,1998Cafiero,2003Hooyberghs,2006Odor,2013Borile}. Many variations of the voter model with quenched disorder have been proposed~\cite{Jedrzejewski2017a}, such as the inclusion of contrarians, defined as agents who adopt a different opinion than that of its neighbors~\cite{2004Galam,2013Masuda}, or zealots, defined as agents who favor~\cite{2003Mobilia,2005Mobilia} or maintain inflexibly~\cite{2007Mobilia,2007Galam} a fixed opinion. This favoring is implemented by including in the dynamics of the zealots an spontaneous transition rate or ``noise'', independent of the copying mechanism, from the disfavored to the favored state. 

Other models include the preference or bias for one of the states in the copying mechanism. This is the case of the {\slshape partisan voter model}~\cite{2010Masuda} in which the population is split into two groups, each one favoring one of the options (Democrats or Republicans in their example), or a model for language competition~\cite{2010Vazquez} in which all agents favor one of the two possible choices, understood then as a difference of prestige between the languages. This particular copying mechanism with preference turns out to be isomorphic to a model of reaction-limited heterogeneous catalysis~\cite{1990Ben,1993Evans} where the bias is the difference in the probabilities of attempting an adsorption of one of two reactive molecular species onto an empty substrate site.

In this work, we consider a variant of the voter model in which a fraction of the population is biased towards one of the two options, while the rest of the population is neutral. Our intention is to determine if the biased community can optimize in some way its connectivity in order to have a maximum influence in the behavior of the whole system. Previous studies on an unbiased majority-like voter model have shown that the opinion held by a minority group can win over that of a larger one provided that it has more internal cohesion (stronger or more connections) than the majority group~\cite{2010Suchecki}. Our setup is very different from previous models with bias in the copying mechanism, such as that of the partisan voter model where each agent displays a bias towards one or another option, or the language model in which all agents favor the same option. 

The structure of the paper is as follows: In Sec. \ref{Sec:model} we present the details of the model. In Sec~\ref{sec:meanfield}, we study the model dynamics on a complete graph using mean-field theory. In Sec.~\ref{sec:random}, we extend the study considering the voters' dynamics on an Erd\H{o}s-R\'enyi random topology, where we assume that the probability of a connection between to sites is independent on whether the voters are biased or not. Finally in Sec.~\ref{Sec:biased}, we extend our studies to the case where voters lie on two Erd\H{o}s-R\'enyi (ER) networks of distinct characteristics, i.e. distinct average degrees, depending on their biased/unbiased type. For this latter scenario, we identify a strategy by which one could change the underlying topology of the network, while preserving the total average degree, in such a way as to achieve a reduction of the consensus time. We show that this is the case when the average degree of the network of interactions among only biased individuals is increased at the cost of decreasing the average degree of the network of interactions of only unbiased voters, while maintaining the average degree of their in-between interactions constant. 

\section{The model} \label{Sec:model}
Let us consider a given lattice of $N$ nodes connected by links. The network is single connected and can not be split in two disjoint ones. Each node $i=1,\dots,N$ represents an agent that holds a binary state ({\slshape opinion}) variable $s_i=\pm1$. In the standard voter model~\cite{1973Clifford,1975Holley} those state variables evolve by an interaction mechanism by which agents copy the opinion of a randomly selected neighbor (those located on connected nodes). We modify these rules by introducing a group of biased agents with preference for one of the states. We consider that $\NB=\gamma N$ agents are biased, and the remaining $\NU=(1-\gamma)N$ are unbiased. Bias is introduced as a parameter $v\in[-1,1]$ which alters the probabilities of biased agents to copy the state of a neighbor. As in the standard voter model dynamics, we select one agent at random (node $i$). Next, one of its neighbors (node $j$) is selected also randomly. If $s_i=s_j$, nothing happens. Else, depending on whether $i$ is biased or unbiased and the state of agent $j$, the following scenarios are considered:
\begin{itemize}
	\item If $i$ is unbiased then $i$ copies $j$'s state with probability $1/2$.
	\item If $i$ is biased: 
	\begin{itemize}
	\item if $s_i=+1$ and $s_j=-1$ then $i$ copies $j$'s state with probability $\dfrac{1-v}{2}$;
	\item if $s_i=-1$ and $s_j=+1$ then $i$ copies $j$'s state with probability $\dfrac{1+v}{2}$.
	\end{itemize}
\end{itemize}
$N$ of these node selections constitute one Monte Carlo step. Under these rules, the preferred state is $s_i=+1$ (resp. $s_i=-1$), $\forall i$ if $v>0$ (resp. $v<0$). The case $v=\pm1$ of extreme preference for one of the two states will not be considered here as it bears some similarity with that of zealot agents, those that never change their opinion, considered elsewhere~\cite{2018Khalil}. Note that the probability for a biased node to copy the state of a neighbor is independent of whether that neighbor is biased or unbiased. The usual voter model is recovered either for $\gamma=0$ (no presence of biased agents) of for $v=0$ when all the nodes follow the standard voter model dynamics with the modification that a neighbor's state is copied only with probability $1/2$. This modification is irrelevant as its only effect is to rescale time (as measured by the number of Monte Carlo steps) by a factor of $2$. As a result of the dynamical rules the system might enter an {\slshape absorbing configuration} where no agent can change its state and no further evolution is possible. With the rules considered here, the only possible absorbing configurations are {\slshape consensus} situations, where all agents hold the same state, either $+1$ or $-1$. The standard voter model also accounts in some situations for dynamical steady states where a macroscopic fraction of agents (but not all of them) hold a particular opinion during a long period of time until a finite size fluctuation takes the system to one of the absorbing configurations. In the biased voter model the symmetry between $+1$ and $-1$ sattes is broken, so that the ensemble average $\langle s_i\rangle$ is no longer a conserved quantity and the system tends to reach the preferred absorbing configuration by its intrinsic dynamics. Still, finite size fluctuations can lead the system to the non-preferred absorbing configuration. 

\section{Mean-field approximation}\label{sec:meanfield}
We present now an analytical treatment of the biased-voter model based on a mean-field type approximation valid for an all-to-all (or complete graph) configuration, where each agent is connected to all other agents. Let $\sigma=\frac{1}{2N}\sum_{i=1}^N(s_i+1)$ be the fraction of network nodes in state $+1$. One can treat the problem as that of a random walk where the variable $\sigma\in[0,1]$ can increase or decrease due to the dynamics. At each time step a randomly selected node $i$ can change its state depending on the state of the randomly selected neighbor $j$. If the change does occur, then the fraction $\sigma$ increases or decreases by an amount $\Delta_\sigma=1/N$. We denote by $R^+(\sigma)$ and $R^-(\sigma)$, respectively, the transition probabilities $\text{Prob}[\sigma \rightarrow \sigma +\Delta_\sigma]$, $\text{Prob}[\sigma \rightarrow \sigma -\Delta_\sigma]$. 
Using the rules of the process described in Section \ref{Sec:model} we can write:
\begin{eqnarray}
R^+(\sigma)&=&\text{Prob}(s_i=-1,s_j=1)\left[\text{Prob}(i \text{ no bias})\frac{1}{2}+\text{Prob}(i\text{ bias})\frac{1+v}{2}\right]\nonumber\\
&=&\text{Prob}(s_j=1|s_i=-1)\text{Prob}(s_i=-1)\left[(1-\gamma)\frac{1}{2}+\gamma\frac{1+v}{2}\right]\nonumber\\
&=&\sigma(1-\sigma)\frac{1+\gamma v}{2}.\label{r_cg}
\end{eqnarray}
Here we have used $\text{Prob}(s_i=-1)=1-\sigma$ and the mean-field approximation: $\text{Prob}(s_j=+1|s_i=-1)=\text{Prob}(s_j=+1)=\sigma$, equivalent to assuming that the state of an agent is independent of that of its neighbors. This implies that the density $\rho$ of active links, those connecting nodes with different opinions, reads as $\rho=\text{Prob}(s_i=-s_j)=2\sigma(1-\sigma)$. Furthermore, we have assumed that the label of ``biased'' or ``unbiased'' of agent $i$ is independent on its state value $s_i$.

Similarly, we derive
\begin{eqnarray}
R^-(\sigma)&=&\text{Prob}(s_i=1,s_j=-1)\left[\text{Prob}(i \text{ no bias})\frac{1}{2}+\text{Prob}(i\text{ bias})\frac{1-v}{2}\right]\nonumber\\
&=&\text{Prob}(s_j=-1|s_i=1)\text{Prob}(s_i=+1)\left[(1-\gamma)\frac{1}{2}+\gamma\frac{1-v}{2}\right]\nonumber\\
&=&\sigma(1-\sigma)\frac{1-\gamma v}{2}. \label{l_cg}
\end{eqnarray}

Once these transition probabilities $R^+(\sigma),\,R^-(\sigma)$ have been derived, one can use the standard machinery of random walk theory~\cite{2001Redner,2010Jacobs} to compute several quantities of interest. We focus in this paper on the fixation probability $P_1$ (probability that all agents eventually reach consensus on the state $+1$), the average time $\tau$ to reach any absorbing state and the average times $\tau_1,\,\tau_{-1}$ to reach consensus on states $+1$ and $-1$, respectively. Without lack of generality, we assume henceforth that $v>0$ such that the preferred state is $+1$.

\subsection{Absorbing state}

The magnetization $m$ is defined in terms of the fraction $\sigma$ as $m=2\sigma-1$. In a single time step $\Delta_t=1/N$ (in units of Monte Carlo steps) $m$ can vary by an amount $\pm\Delta_m=\pm 2\Delta_\sigma=\pm2/N$. The evolution equation for the probability $P(m, t)$ of finding a magnetization $m$ at time $t$ follows from the basic rules of the process as
\begin{eqnarray}
P(m,t+\Delta_t)&=&R^{-}(m+\Delta_m)P(m+\Delta_m,t)\\
&+& R^{+}(m-\Delta_m)P(m-\Delta_m,t)\nonumber\\
&+& [1-R^{+}(m)-R^{-}(m)]P(m,t).\nonumber
\end{eqnarray} 
where $R^\pm(m)$ are the transition probabilities Eqs.(\ref{r_cg},\ref{l_cg}) written in terms of the $m$ variable. Upon Taylor expanding up to order $\Delta_t$ in time and up to second order in $\Delta_m$ or, equivalently, taking the continuous limit $1/N\rightarrow0$ and keeping only terms up to order $1/N$, the time evolution of $P(m,t)$ is given by
\begin{eqnarray}
\frac{\partial P(m,t)}{\partial t}&=&\frac{\Delta_m}{\Delta_t}\frac{\partial[(R^{-}(m)-R^{+}(m))P(m,t)]}{\partial m}\nonumber\\&&+\frac{(\Delta_m)^2}{2 \Delta_t}\frac{\partial^2 [(R^{-}(m)+R^{+}(m))P(m,t)]}{\partial m^2},
\end{eqnarray}
which, after replacing the expressions for $R^\pm(m)$, $\Delta_m$ and $\Delta_t$, becomes 
\begin{eqnarray}
\frac{\partial P\left(m,t\right)}{\partial t}=-\frac{\gamma v}{2}\frac{\partial \left[\left(1-m^{2}\right)P\left(m,t\right)\right]}{\partial m}+\frac{1}{2N}\frac{\partial^{2}\left[\left(1-m^{2}\right)P\left(m,t\right)\right]}{\partial m^{2}}.\label{Fokker-Planck}
\end{eqnarray}
For the case $\gamma v=0$ this reproduces the results of the standard voter model~\cite{1990Ben,2003Slanina} with the already mentioned rescaling of the time by a factor of $2$.

Eq.\eqref{Fokker-Planck} is a Fokker-Planck equation with state-dependent drift $F(m)=\frac{\gamma v}{2}(1-m^2)$ and diffusion coefficient $D\left(m\right)=\frac{1-m^{2}}{N}$, hence the evolution of the magnetization can be viewed as the motion of a random walk moving in a medium that is increasingly ``sticky" near the extremities of the absorbing interval. When $m=\pm1$ the walk stops, independently of whether there is bias or not. In principle, the steady-state solution of the Fokker-Planck Eq.\eqref{Fokker-Planck} would be $P_\text{st}\left(m\right)=Z^{-1}\frac{e^{\gamma vN m}}{1-m^{2}}$. However, the normalization constant is $Z=\int_{-1}^{1}\frac{e^{\gamma vNm}}{1-m^{2}}dm=\infty$,
as the integral diverges in both limits $m=\pm 1$. This indicates that the only absorbing state for a finite size system is consensus to either of these two states. Given that $\gamma v>0$, we conclude that in the thermodynamic limit it is $P_\text{st}(m)=\delta(m-1)$, being $\delta(\cdot)$ the Dirac-delta function, and the stationary consensus state will be the preferred one $m=1$. To verify this, we derive from the Fokker-Planck Eq.(\ref{Fokker-Planck}) the following equation of motion for the average magnetization $\langle m\rangle$
\begin{eqnarray}
\frac{\partial \langle m\rangle}{\partial t}=\frac{\gamma v}{2}(1-\langle m^2\rangle).\label{magnetization_complete}
\end{eqnarray}
Neglecting fluctuations, $\langle m^2\rangle\approx\langle m\rangle^2$, the solution for an initial condition $\langle m(0)\rangle=0$ is $\langle m(t)\rangle= \tanh(\frac{\gamma v}{2} t)$ as shown in~\cite{2010Vazquez} for the case $\gamma=1$. This describes a monotonic evolution to the stationary state in a characteristic time scale $1/(\gamma v)$. In the following we consider the effect of finite size fluctuations that can lead the system to the non-preferred absorbing state.

\subsection{Fixation probability, $P_1$}
The fixation (or exit) probability $P_1(\sigma)$ is defined as the probability that a finite system with an initial fraction $\sigma$ reaches a consensus to the preferred state $+1$ in a finite number of steps~\cite{Krapivsky:2010}. It can be expressed as the probability of making one of the transitions $\sigma \to \sigma-\Delta_\sigma,\,\sigma,\,\sigma+\Delta_\sigma$ multiplied by the exit probability from these intermediate points:
\begin{eqnarray}
P_1(\sigma)&=&R^+(\sigma)P_1(\sigma+\Delta_\sigma)+R^-(\sigma)P_1(\sigma-\Delta_\sigma)\nonumber \\
&+&[1-R^+(\sigma)-R^-(\sigma)]P_1(\sigma),
\label{p1_1}
\end{eqnarray}
with the boundary conditions $P_1(0)=0$, $P_1(1)=1$. 
Introducing the notation $P_1(n)=P_1(\sigma=\frac{n}{N})$, and using Eqs.(\ref{r_cg},\ref{l_cg}), Eq.(\ref{p1_1}) is rewritten as a recurrence equation
\begin{equation}
P_1(n)=\frac{1+\gamma v}{2}P_1(n+1)+\frac{1-\gamma v}{2}P_1(n-1).
\label{p1_3}
\end{equation}
The solution of this equation satisfying the aforementioned boundary conditions is:
\begin{equation}
P_1(n)=\frac{1-a^n}{1-a^N},\quad a=\dfrac{1-\gamma v}{1+\gamma v},
\label{p1_4}
\end{equation}
which for random initial conditions $n=\frac{N}{2}$ takes the form
\begin{equation}
P_1(n=N/2)=\frac{1}{1+a^{N/2}}.
\label{p1_5}
\end{equation}
If we approximate $\ln a=-2\gamma v+O(\gamma v)^3$, at this order in the expansion of the logarithm, we obtain $a^{N/2}\approx \exp(-\gamma v N)$, which leads to
\begin{equation}
P_1(n=N/2)=\frac{1}{1+\exp(-\beta/2)}, \,\beta=2\gamma v N.
\label{p1_6}
\end{equation} 
Note that the exit probability in this approximation is only a function of the product $\gamma v N=v\NB$. 
It turns out that the approximation is very good and the maximum absolute difference between Eqs.(\ref{p1_5}) and (\ref{p1_6}) is always smaller than $0.42N^{-2}$ for all values of $\gamma v$, such that for the system size $N=1000$ used in most numerical simulations of this paper, the error of the approximation is less then $4.2\times 10^{-7}$. 

\begin{figure}[!h]
\includegraphics[width = 0.5\columnwidth, keepaspectratio = true]{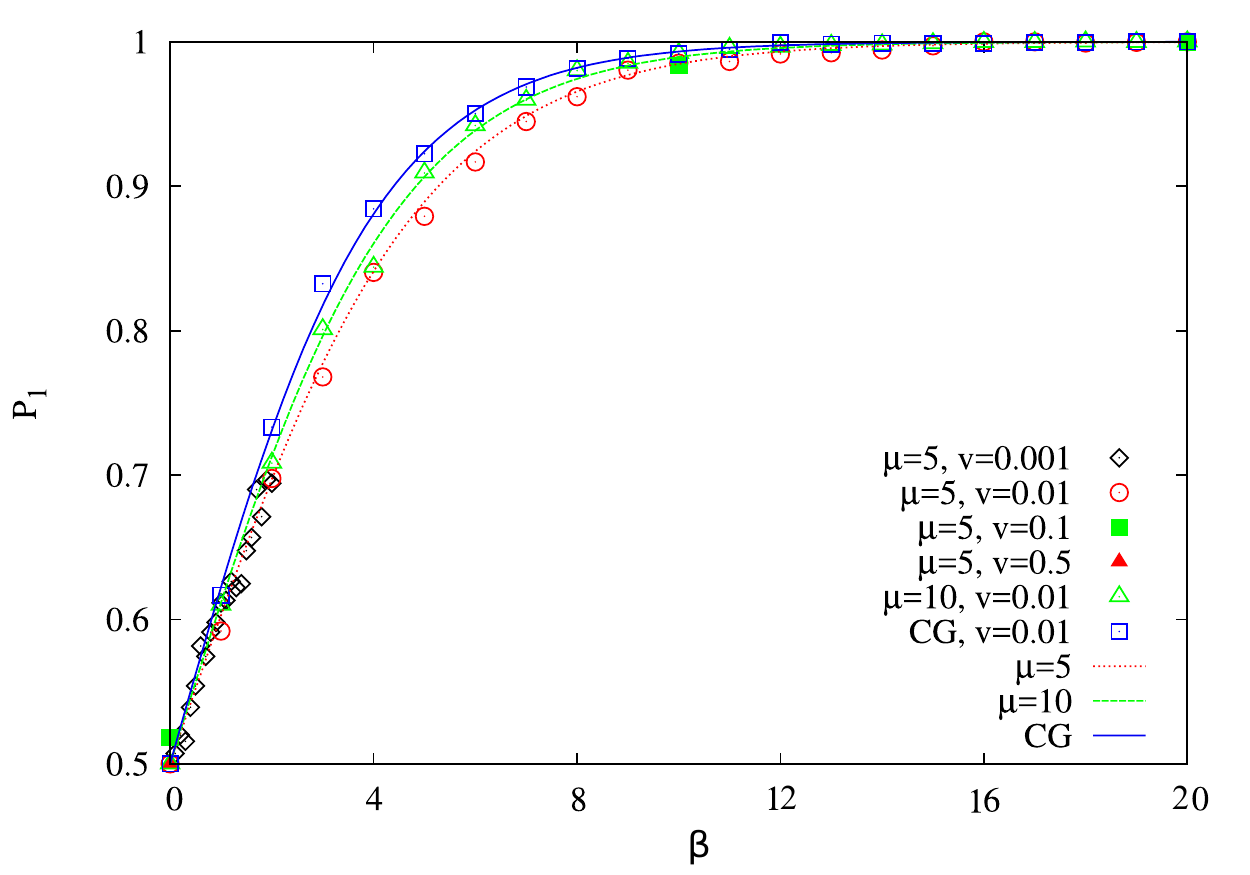}
\includegraphics[width = 0.5\columnwidth, keepaspectratio = true]{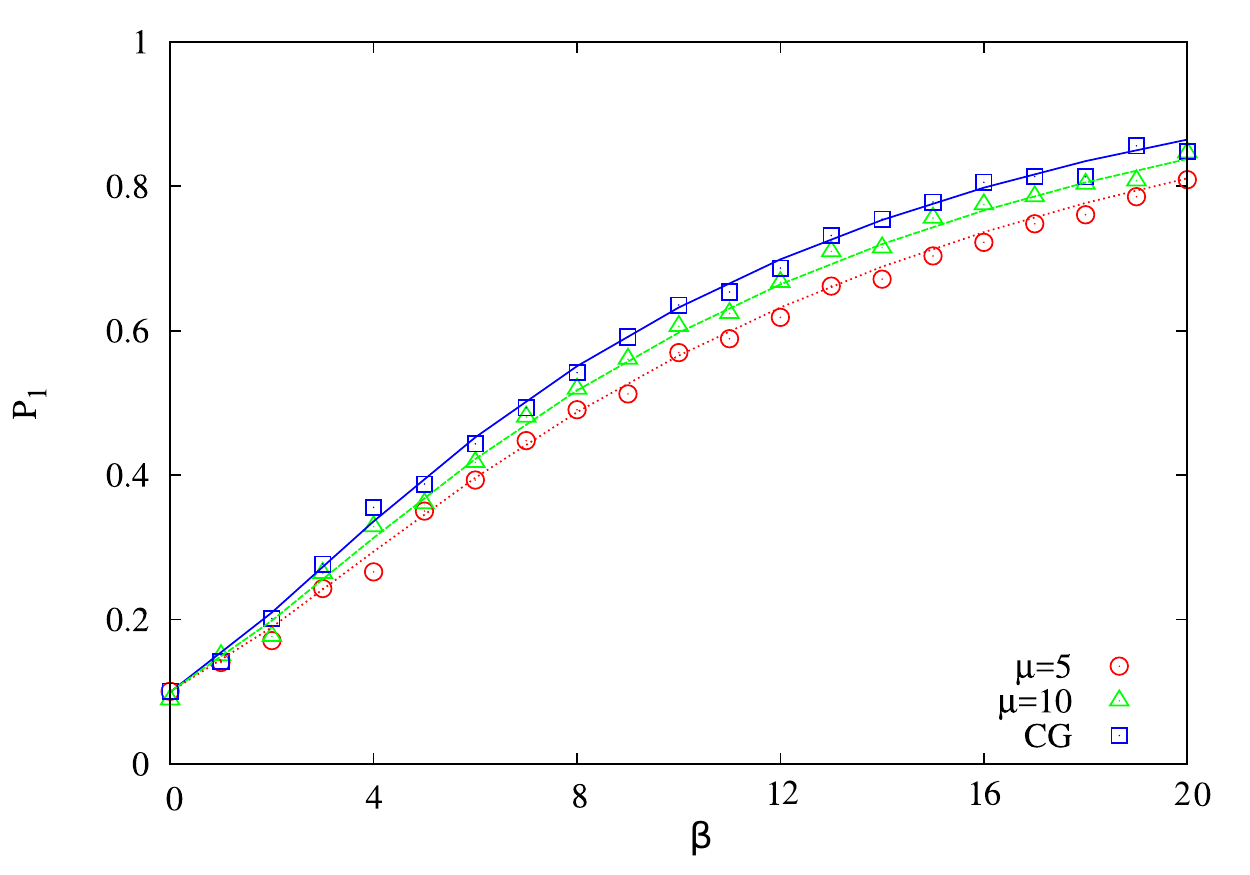}
\caption{Plot of the fixation probability (probability to reach the preferred state $s=+1$), $P_1(\sigma, \beta)$, as a function of the scaling variable $\beta=2\gamma v N$ for two different values of the initial condition: $\sigma=0.5$ (left panel) and $\sigma=0.1$ (right panel), using values $\gamma=0.1$ and $N=1000$. The continuous (blue) lines are the analytical result, Eq.(\ref{p1_9}), and the open square ($\square$) symbols denote the results of computer simulations in a complete graph (CG) for $v=0.01$. We also plot the fixation probability for two random Erd\H{o}s-R\'eny networks with average connectivity $\mu=5$ and $\mu=10$ and several values of $v$ as indicated in the legend of the left panel, and $v=0.01$ in the case of the right panel. The theoretical predictions lines, dashed (green) for $\mu=10$ and dotted (red) for $\mu=5$, come from the same expression Eq.(\ref{p1_9}) replacing $\beta$ by $\beta_\mu=2\gamma v \dfrac{\mu}{\mu+1}N$ as explained in section \ref{sub:P1_networks}, while the different symbols are the results of numerical simulations.}
\label{fig_p1_cg_er}
\end{figure}

It is worth noting that the approximate solution Eq.(\ref{p1_6}) can also be obtained from a continuous version of Eq.(\ref{p1_1}), obtained by expanding to second order in $\Delta_\sigma$:
\begin{equation}
\frac{1}{2}\Delta_\sigma^2\left(R^++R^-\right)\frac{d^2 P_1}{d\sigma^2}+\Delta_\sigma\left(R^+-R^-\right)\frac{dP_1}{d\sigma}=0,
\label{p1_7}
\end{equation}
the so-called backward Kolmogorov equation for the exit probability. After replacing Eqs.(\ref{r_cg},\ref{l_cg}) and $\Delta_\sigma=1/N$, we obtain
\begin{equation}
\frac{d^2 P_1}{d\sigma^2}+\beta\frac{dP_1}{d\sigma}=0,
\label{p1_8}
\end{equation}
whose solution for boundary conditions $P_1(0)=0$ and $P_1(1)=1$ is~\footnote{\label{Note1}The general solution of $$z''(\sigma )+\beta z'(\sigma )=g(\sigma )$$ is $$z(\sigma )=C_2+C_1e^{-\beta \sigma }+\int^\sigma d\sigma' \int^{\sigma '}d\sigma''e^{-\beta (\sigma'-\sigma'')}g(\sigma''),$$ where $C_1,\,C_2$ are integration constants found by fulfilling the adequate boundary conditions.}
\begin{equation}
P_1(\sigma;\beta)=\frac{1-e^{-\sigma\beta}}{1-e^{-\beta}}.
\label{p1_9}
\end{equation}
Setting the initial fraction $\sigma=\frac{1}{2}$ we recover Eq.(\ref{p1_6}). Furthermore, when $\gamma v =0$ we recover the well known formula for the standard voter model $P_1(\sigma)=\sigma$.
As shown in Fig.~\ref{fig_p1_cg_er} the analytical expression Eq. (\ref{p1_6}) agrees well with the data coming from numerical simulations of the model on a complete graph. The figure also shows results corresponding to the random (Erd\H{o}s-R\'eny) network for different values of the average connectivity $\mu$ that will be analyzed in detail in Sec.\ref{sec:random}. 

\begin{figure}[h!]
\centering
\includegraphics[width = 0.5\columnwidth, keepaspectratio = true]{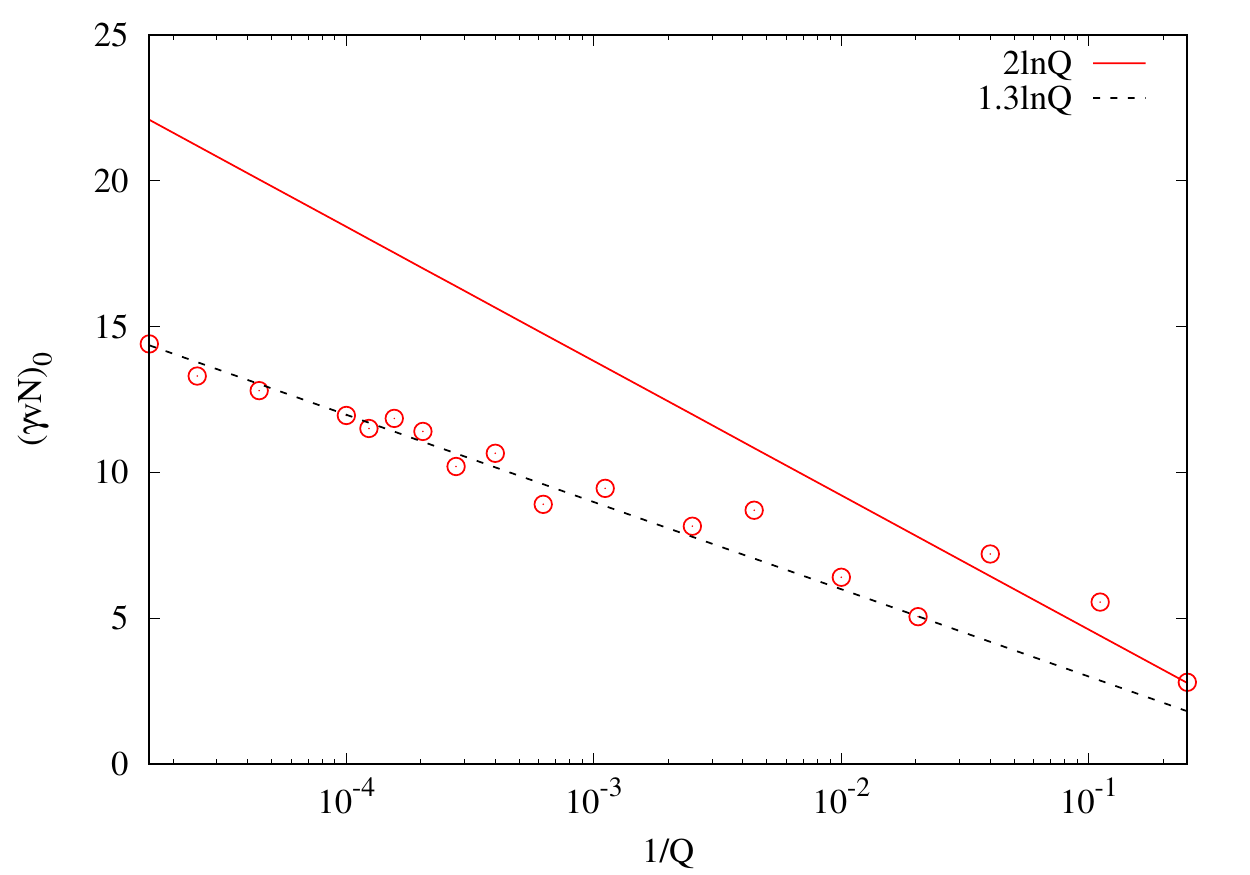}
\caption{In this figure we plot the threshold value $(\gamma v N)_0$ as a function of the number of runs $Q$ as obtained from numerical simulations of the biased-voter model in a complete-graph (all-to-all connectivity between the agents). As explained in the main text, $(\gamma v N)_0$ is defined as the value of $\gamma v N$ above which none of the $Q$ runs is observed to end in the non-preferred state $s=-1$. The solid line is the approximation $2\ln(Q)$ as given by Eq.(\ref{gvca}), while the dashed line is a fit to the form $1.3\ln Q$.}
\label{realizations}
\end{figure}

A consequence of Eq.(\ref{p1_9}) is that, as the system size increases, the probability $1-P_1(\sigma;\beta)$ of reaching consensus in the non-preferred opinion decreases and, eventually, tends to zero as $N$ tends to infinity for any non-zero value of the product $\gamma v$. For a finite system size, however, there is a finite probability to reach the non-preferred state, which means that, in principle, it is possible to observe some realizations of the dynamics leading to consensus for this non-preferred state. A word of caution is relevant here: due to the smallness of this probability for large system sizes, a large number $Q$ of realizations is needed to observe a consensus in the non-preferred state in a numerical simulation of the process. Alternatively, for a given number or realizations $Q$ there will be a value of $(\gamma v N)_0$ above which the probability to reach the non-preferred state is smaller that $1/Q$ and no consensus to the non-preferred state will be likely to be observed in the numerical simulation, leading to the wrong conclusion that order in this non-preferred state is never possible if $\gamma v N >(\gamma v N)_0$. This value of $(\gamma v N)_0$ can be estimated by noting that the probability that at least one of the $Q$ runs ends in the non-preferred state ($s_i=-1,\,\forall i$, remember that we assume $\gamma v>0$) is $1-P_1^Q$ and we demand this probability to be of the order the inverse of the number of runs $1-P_1^Q\sim 1/Q$. Using Eq.(\ref{p1_6}), we arrive at the condition
\begin{equation}
(\gamma v N)_0 \sim -\ln\left((1-\frac{1}{Q})^{-1/Q}-1\right)=2\ln Q +O(1/Q)\label{gvca}
\end{equation}
If $\gamma v N> (\gamma v N)_0$ then no runs will be typically observed to reach the non-preferred state in the $Q$ runs of the simulation. In practice, we observe that the threshold value $(\gamma v N)_0$ scales roughly as $1.3\ln Q$, see Fig.~\ref{realizations}.

\subsection{Time to reach consensus, $\tau$}

Let $\tau(\sigma)$ be the average time to reach any absorbing state (all agents taking the same value, $s_i=s,\,\forall i$ with either $s=+1$ or $s=-1$) starting from an initial fraction $\sigma$. As before, we consider the transitions $\sigma \to \sigma-\Delta_\sigma, \sigma,\sigma+\Delta_\sigma$ and relate $\tau(\sigma)$ to the average times from these intermediate points,
\begin{eqnarray}
\tau(\sigma)&=&R^+(\sigma)\left[\tau(\sigma+\Delta_\sigma)+\Delta_t\right]+ R^-(\sigma)\left[\tau(\sigma-\Delta_\sigma)+\Delta_t\right]\nonumber\\
&&+\left[1-R^+(\sigma)-R^-(\sigma)\right]\left[\tau(\sigma)+\Delta_t\right]
\label{ttrc_1},
\end{eqnarray}
and boundary conditions $\tau(0)=\tau(1)=0$. Replacing $\sigma=n/N$, introducing the notation $\tau(n)=\tau(\sigma=n/N)$, and using Eqs.(\ref{r_cg},\ref{l_cg}) we arrive at the difference equation:
\begin{equation}
\frac{1+\gamma v}{2}\tau(n+1)+\frac{1-\gamma v}{2}\tau(n-1)=\tau(n)-\frac{1}{n}-\frac{1}{N-n},
\label{ttrc_d2}
\end{equation}
to be solved with the aforementioned boundary conditions. 
Due to the linearity of this difference equation and the symmetry of the process under the change $n\to N-n$ and $v\to -v$, the solution can be written as
\begin{equation}\label{ttrc_d7}
\tau(n)=T_1(n;\gamma v)+T_1(N-n;-\gamma v),
\end{equation}
where $T_1(n;\gamma)$ is the solution satisfying the boundary conditions $T_{1}(0)=T_{1}(N)=0$ of the following difference equation:
\begin{eqnarray}
\frac{1+\gamma v}{2}T_1(n+1)+\frac{1-\gamma v}{2}T_1(n-1)&=&T_1(n)-\frac{1}{n}\label{ttrc_d4}.
\end{eqnarray}
As explained in the Appendix, the solution can be written in terms of the harmonic function $H_n$ and the function $f(n,a)$ defined in Eqs.(\ref{eq:hn},\ref{eq:fna}) as:
\begin{eqnarray}\label{ttrc_d20}
T_1(n;a)=\frac{1+a}{1-a}\biggl[\dfrac{1-a^n}{1-a^N}\left[H_{N-1}-f(N,a)\right]-\left[H_{n-1}-f(n,a)\right]\biggr].
\end{eqnarray}

According to the definition Eq.(\ref{p1_4}), when $\gamma v$ is replaced by $-\gamma v$, $a$ becomes $\frac{1}{a}$ and Eq.(\ref{ttrc_d7}) can be written as 
\begin{equation}
\tau(n;a)=T_1(n;a)+T_1(N-n;\frac{1}{a}).
\label{ttrc_d15}
\end{equation}

\begin{figure}[H]
\centering
\includegraphics[width = 0.5\columnwidth, keepaspectratio = true]{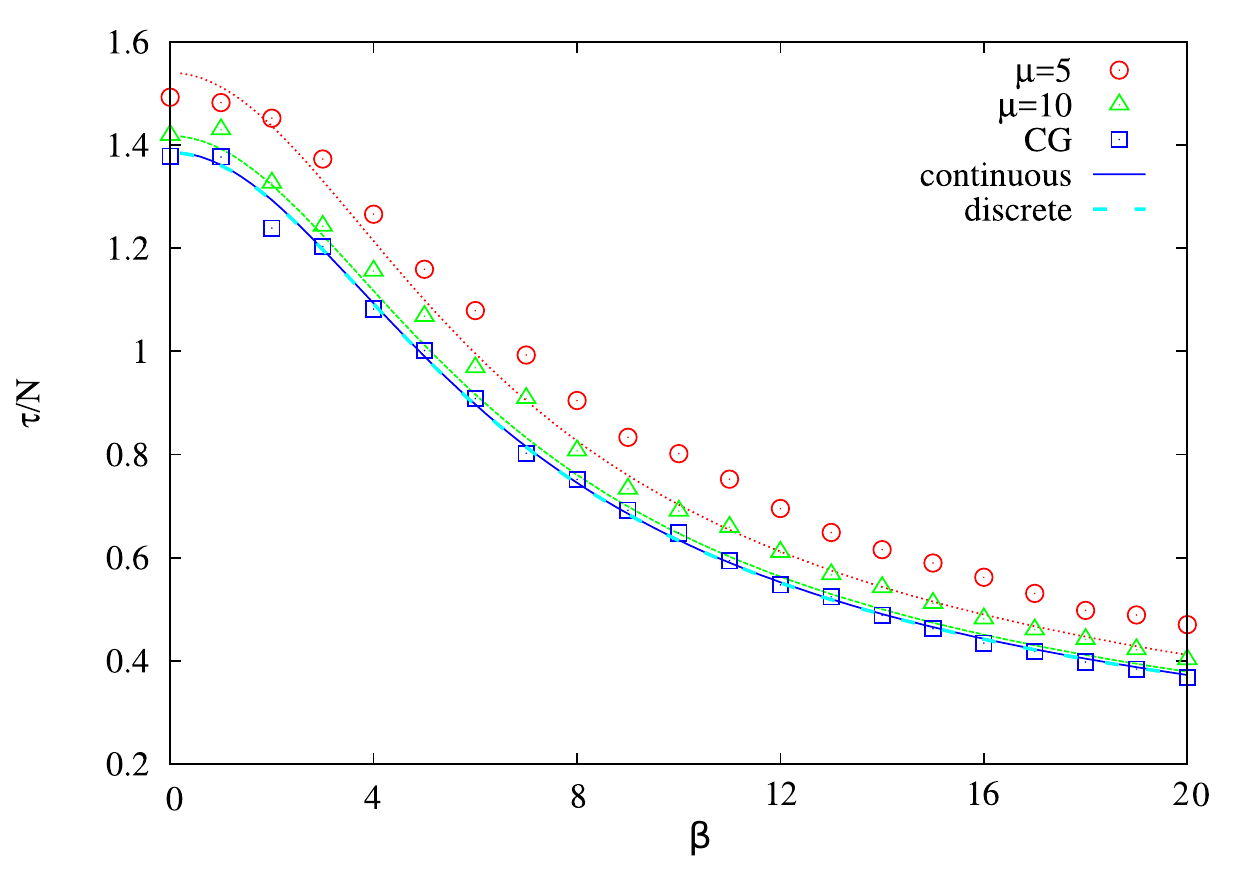}
\label{er_cg_tau}
\caption{Average time $\tau$ to reach an absorbing state, in units of MCS and rescaled by the system size $N$, for an initial condition $\sigma=0.5$ as a function of $\beta=\gamma v N$. At the scale of the figure the analytical predictions of the continuous approach Eq.(\ref{ttrc_6}), continuous (blue) line, and the discrete time approach Eq.(\ref{ttrc_d15}), long-dashed (light blue) line, are indistinguishable. The open square ($\square$) symbols denote the results of computer simulations in a complete graph (CG) for $v=0.01$, $\gamma=0.1$ and $N=1000$. We also plot the results for the random Erd\H{o}s-R\'eny networks with average degree $\mu$. The theoretical prediction lines, dashed (green) for $\mu=5$ and dotted (red) for $\mu=10$ are given by Eq.(\ref{eq:tau_network}), and the different symbols are the results of numerical simulations.}
\label{er_cg_ttrc}
\end{figure}

It is also possible to obtain an approximation to this expression starting directly from the differential equation that follows from the expansion of Eq.(\ref{ttrc_1}) to second order in $\Delta_\sigma$:
\begin{equation}
\frac{1}{2}\frac{\Delta_\sigma^2}{\Delta_t}\left(R^++R^-\right)\frac{d^2\tau}{d\sigma^2}+\frac{\Delta_\sigma}{\Delta_t}\left(R^+-R^-\right)\frac{d\tau}{d\sigma} =-1,
\label{ttrc_2}
\end{equation}
or, replacing the expressions for $R^+,R^-,\Delta_\sigma,\Delta_t$,
\begin{equation}
\frac{d^2\tau}{d\sigma^2}+\beta\frac{d\tau}{d\sigma} =-2N\left(\frac{1}{\sigma}+\frac{1}{1-\sigma}\right).
\label{ttrc_3}
\end{equation}
The solution of this equation with boundary conditions $\tau(0)=\tau(1)=0$ is
\begin{equation}
\tau(\sigma;\beta)= 2N\left[T(\sigma;\beta)+ T(1-\sigma;-\beta)\right],
\label{ttrc_6}
\end{equation}
with $T(\sigma;\beta)$ the solution of
\begin{eqnarray}
\frac{d^2T}{d\sigma^2}+\beta\frac{dT}{d\sigma} =-\frac{1}{\sigma},&&\nonumber\\
T(0)=T(1)=0.\label{ttrc_4}
\end{eqnarray}
given explicitly by (see footnote~\ref{Note1}) 
\begin{eqnarray} \label{ttrc_7c}
T(\sigma;\beta)=\frac{1}{\beta}\biggl[\frac{e^{-\beta\sigma}-1}{e^{\beta}-1}\Ei(\beta)+\frac{1-e^{\beta(1-\sigma)}}{e^\beta-1}(\ln(|\beta |)+\gamma_\text{e})+e^{-\beta\sigma} \Ei(\beta \sigma )-\ln(\sigma)\biggr],
\end{eqnarray}
where $\gamma_{e}\approx0.577$ is the Euler-Mascheroni constant and $\Ei(x)=-\int_{-x}^{\infty} \exp(-z)/zdz$ the exponential integral. Again, the continuous approximation Eq.(\ref{ttrc_6}) and the discrete counterpart Eq.(\ref{ttrc_d15}) are almost indistinguishable for all system sizes and parameter values used in the figures. An example is given in Fig.~\ref{er_cg_tau}, where we compare the results of this theoretical analysis with those of computer simulations on a complete-graph and a random initial condition $\sigma=0.5$. In the same plot we see results corresponding to the random (Erd\"os-R\'eny) network for different values of the average connectivity $\mu$ that we will address in a more detailed theory developed in Sec.~\ref{sec:random}.

\begin{figure}
 \centering
\includegraphics[width = 0.5\columnwidth, keepaspectratio = true]{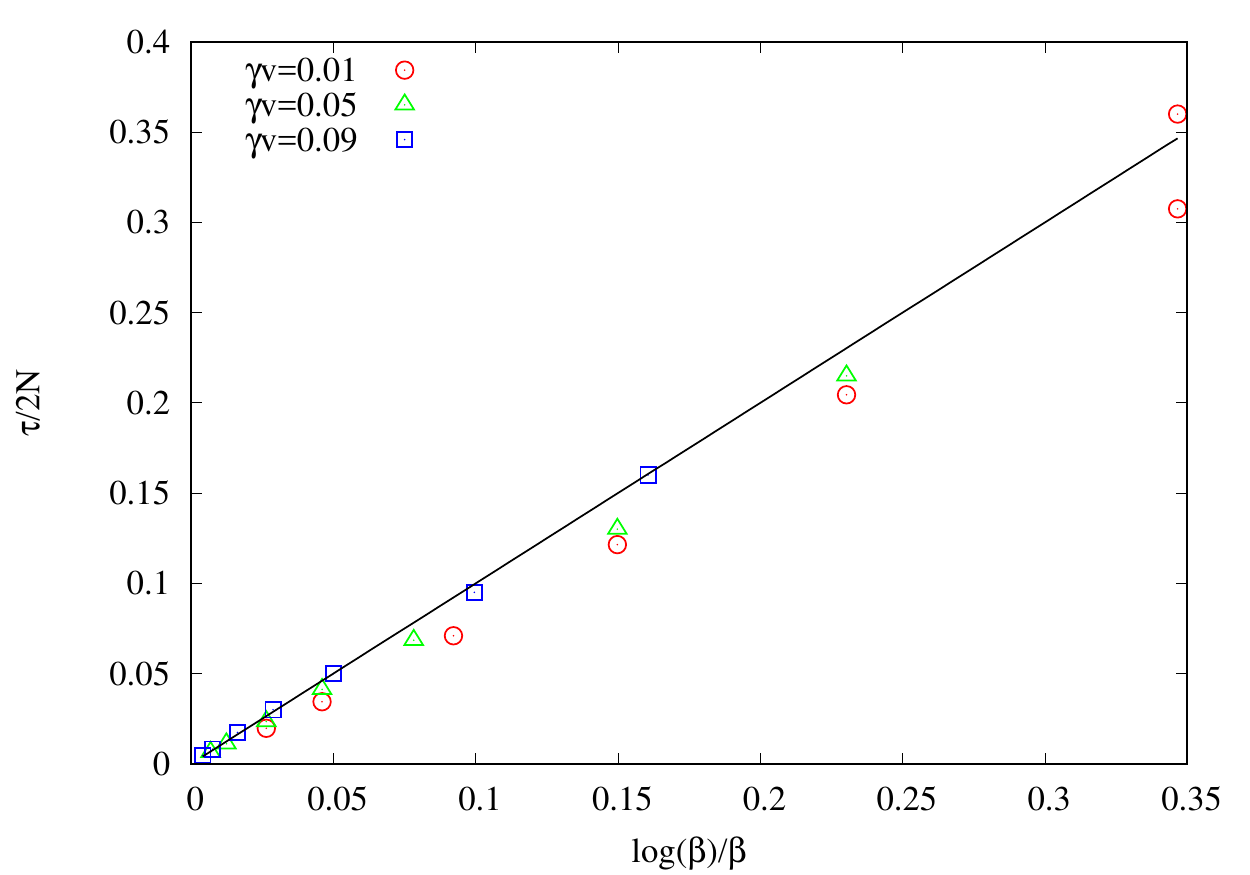}
\caption{We plot the average time $\tau$ to reach an absorbing state, in units of MCS, in order to check the logarithmic dependence with system size $N$ as it follows from the asymptotic expression Eq. (\ref{ttrc1_approx}), written as $\frac{\tau}{2N}\to\frac{\log \beta}{\beta}$, solid line. The symbols correspond to numerical simulations on a complete graph for different values of $\gamma v$, as indicated in the legend. }
\label{fig_tau_cg_erN}
\end{figure}

Using the known asymptotic expansions of the exponential integral $\lim_{x\to \infty} \Ei(x)= e^{x}(x^{-1}+O(x^{-2}))$ we find that:
\begin{equation}
\lim_{\beta\rightarrow\infty}\tau(\sigma=1/2;\beta)\rightarrow 2N\frac{\ln(\beta)+\gamma_\text{e}}{\beta}\to \frac{\ln(N)}{\gamma v},
\label{ttrc1_approx}
\end{equation}
which means that the average time, in units of MCS, to reach consensus starting from $\sigma=1/2$ in the presence of a group of biased agents, $\gamma v\ne0$, scales with the number of agents as $\ln(N)$. A result that is confirmed by the numerical simulations, see Fig.~\ref{fig_tau_cg_erN}. This is to be compared with the limit of no bias $\gamma v\to 0$ which can be obtained directly from Eq.(\ref{ttrc_3}) setting $\beta=0$, or from Eq.(\ref{ttrc_6}) using the expansion $\lim_{x\to 0} \Ei(x)= \ln(|x|)+\gamma_\text{e}+x+O(x^2)$ 
\begin{equation}
\lim_{\beta\rightarrow 0}\tau(\sigma=1/2;\beta)\rightarrow N\ln(4),
\label{ttrc1_approx2}
\end{equation}
a much slower and well known approach to consensus than in the biased case.

\subsection{Time to reach preferred state, $\tau_1$}

\begin{figure}[!h]
\subfigure[$\beta=1$]{
\includegraphics[width = 0.5\columnwidth, keepaspectratio = true]{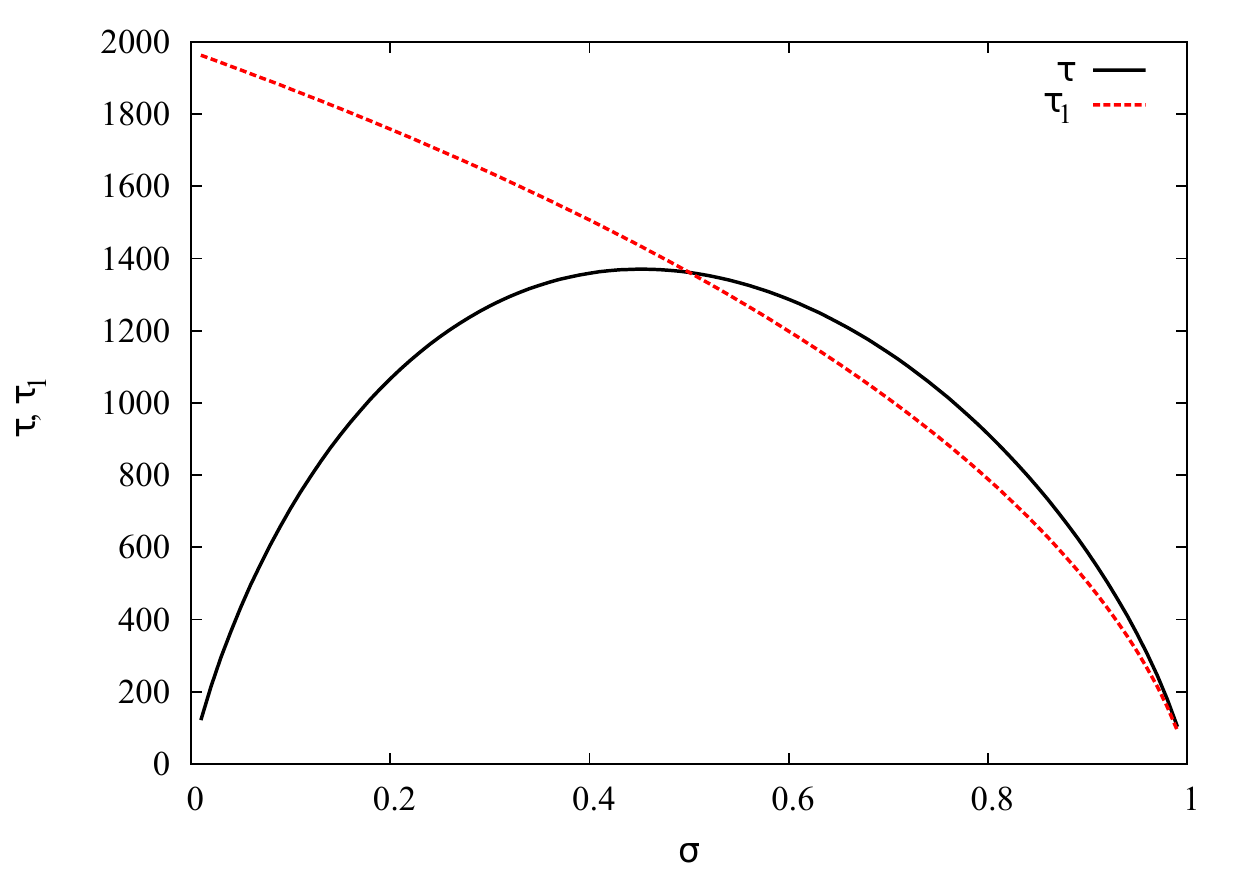}
\label{fig_tau_b1}
}
\subfigure[$\beta=10$]{
\includegraphics[width = 0.5\columnwidth, keepaspectratio = true]{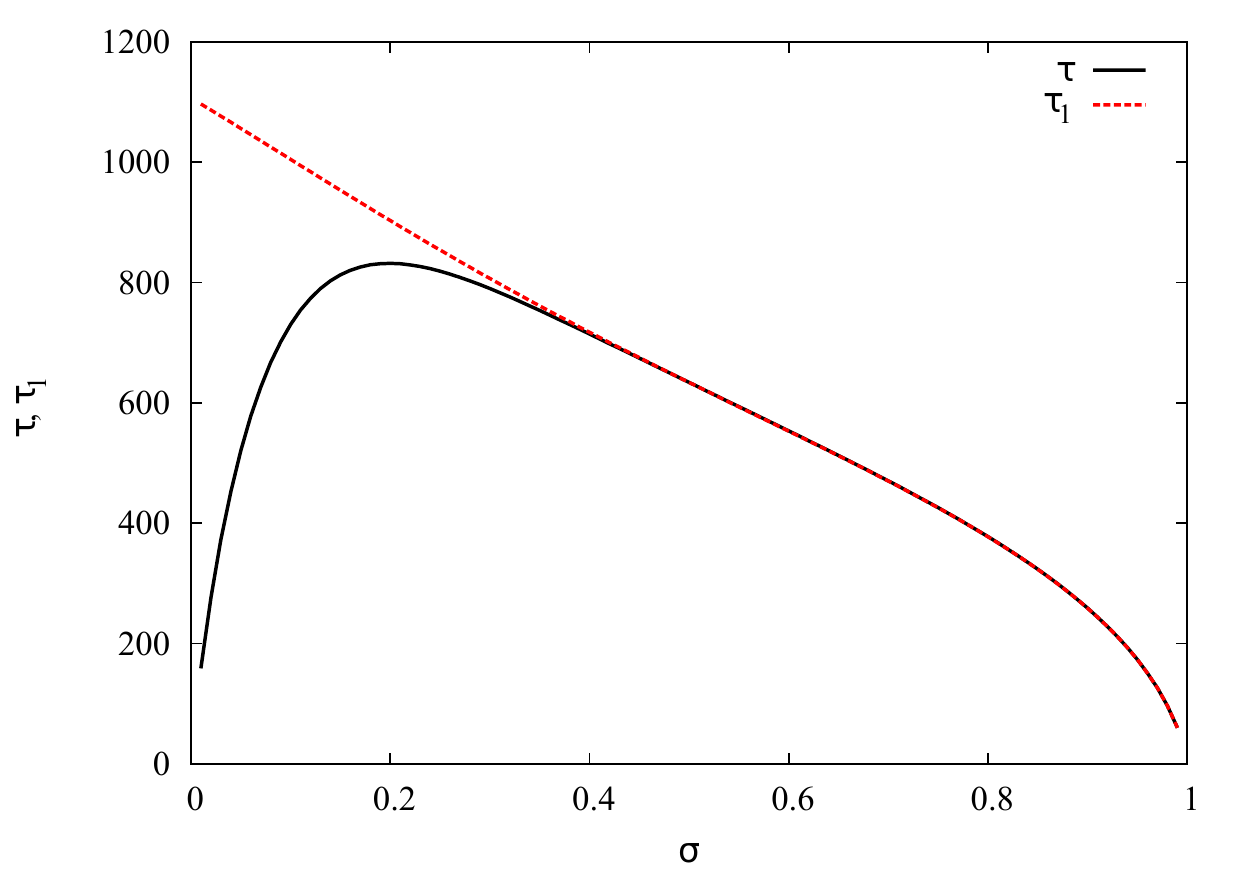}
\label{fig_tau_b10}
}
\caption{Plot of the average time to reach consensus in any state starting from an initial fraction $\sigma$ of agents in the preferred state, $\tau(\sigma)$, as given by Eq.(\ref{ttrc_6}) continuous (black) line, and the average time to reach consensus in the preferred state, $\tau_1(\sigma)$, given by Eq.(\ref{ttrc1_8}) dotted (red) line for $\beta=1$ (left panel), and $\beta=10$ (right panel). Note that both times, and hence also the time $\tau_{-1}$ to reach consensus in the non-preferred state, see Eq.(\ref{ttrc_9}), coincide for $\sigma=1/2$. Parameters values: $\gamma=0.1,v=0.01, N=1000$.}
\label{fig_tau_b}
\end{figure}

The average time $\tau_1(\sigma)$ to reach the preferred absorbing state (for $\gamma v>0$ the preferred state is $+1$) starting from an initial fraction $\sigma$ satisfies a recurrence relation:
\begin{eqnarray}
P_1(\sigma)\tau_1(\sigma)&=&R^+(\sigma)\left[P_1(\sigma+\Delta_\sigma)\tau_1(\sigma+\Delta_\sigma)+P_1(\sigma)\Delta_t\right]\nonumber \\ &+& R^-(\sigma)\left[P_1(\sigma-\Delta_\sigma)\tau_1(\sigma-\Delta_\sigma)+P_1(\sigma)\Delta_t\right] \label{ttrc1_1}\nonumber\\ 
&+&\left[1-R^+(\sigma)-R^-(\sigma)\right]\left[P_1(\sigma)\tau_1(\sigma)+P_1(\sigma)\Delta_t\right].
\end{eqnarray}
We do not solve this recurrence relation, but proceed directly to the continuous limit approach, in view of its accuracy. Expanding $P_1\tau_1$ to second order in $\Delta_\sigma=\Delta_t=1/N$, Eq.(\ref{ttrc1_1}) becomes
\begin{equation}
\frac{1}{2N}\left(R^++R^-\right)\frac{d^2(P_1\tau_1)}{d\sigma^2}+\left(R^+-R^-\right)\frac{d(P_1\tau_1)}{d\sigma} =-P_1.
\label{ttrc1_2}
\end{equation}
Replacing $P_1(\sigma)$ from Eq.(\ref{p1_9}) in the right-hand-side, we obtain
\begin{equation}
\frac{d^2(P_1\tau_1)}{d\sigma^2}+\beta\frac{d(P_1\tau_1)}{d\sigma} =-\frac{2N}{1-e^{-\beta}}\left(\frac{1}{\sigma}+\frac{1}{1-\sigma}-\frac{e^{-\beta\sigma}}{\sigma}-\frac{e^{-\beta\sigma}}{1-\sigma}\right)
\label{ttrc1_3}
\end{equation}
with boundary conditions $\tau_1(1)P_1(1)=\tau_1(0)P_1(0)=0$. To solve this equation we note that the solution of
\begin{eqnarray}
\frac{d^2\hat T}{d\sigma^2}+\beta\frac{d\hat T}{d\sigma} =-\frac{e^{-\beta \sigma}}{\sigma},&&\nonumber\\
\hat T(0)=\hat T(1)=0,&&
\label{hatt}
\end{eqnarray} 
is $\hat T(\sigma;\beta)=e^{-\beta\sigma}T(\sigma,-\beta)$, with $T(\sigma;\beta)$ as given by Eq.(\ref{ttrc_7c}). The solution of (\ref{ttrc1_3}) is hence
\begin{eqnarray}
\tau_1(\sigma;\beta)=\frac{2N}{1-e^{-\beta\sigma}}\bigl[T(\sigma;\beta)+T(1-\sigma,-\beta)\nonumber \\ 
 -e^{-\beta\sigma} T(\sigma,-\beta)-e^{-\beta\sigma} T(1-\sigma;\beta)\bigr].\label{ttrc1_8}
\end{eqnarray}
In Fig.~\ref{fig_tau_b} we plot the times $\tau(\sigma;\beta)$ and $\tau_1(\sigma;\beta)$ as given by Eqs.(\ref{ttrc_6}, \ref{ttrc1_8}), as a function of the initial value $\sigma$ for two different values of $\beta$. For the particular case of random initial conditions $\sigma=1/2$, it turns out that $\tau(\sigma=1/2;\beta)=\tau_1(\sigma=1/2;\beta)$. When the initial fraction $\sigma$ of $+1$ agents is smaller than $1/2$, then even in the presence of bias it is $\tau_1>\tau$. In the opposite case, when we run the dynamics starting with more than half of the agents in the preferred state and small values of $\beta$ (see Fig.~\ref{fig_tau_b1}), we observe that $\tau_1<\tau$. A similar relation between $\tau$ and $\tau_1$ holds for the standard voter model~\cite{2008Sood}. As showed in Fig.~\ref{fig_tau_b10}, for a large bias parameter $\beta=2\gamma v N$ both times converge to the same value, a result which is a consequence of the very small probability to reach the non-preferred state. 

\begin{figure}[H]
\subfigure[$\tau_1$]{
\includegraphics[width = 0.5\columnwidth, keepaspectratio = true]{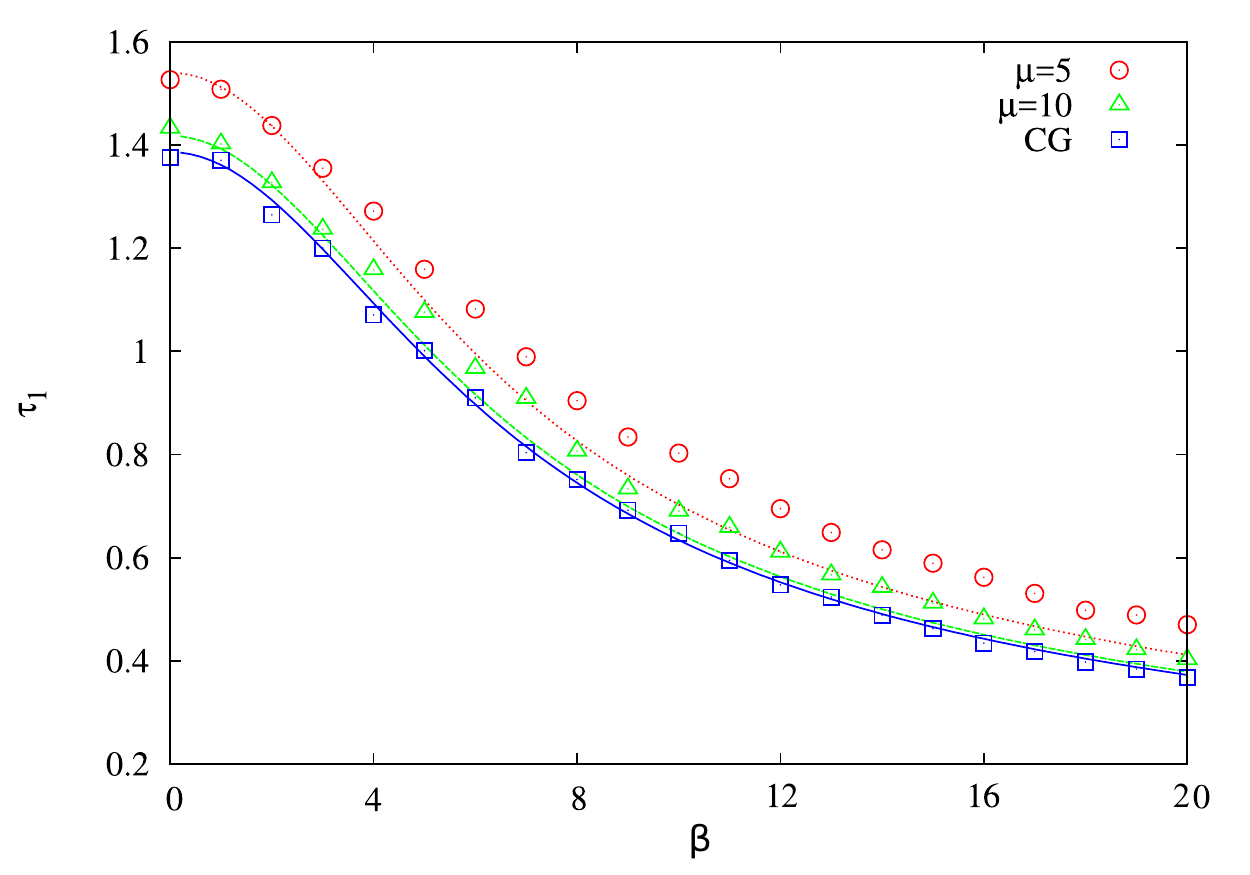}
\label{er_cg_tau1}
}
\subfigure[$\tau_{-1}$]{
\includegraphics[width = 0.5\columnwidth, keepaspectratio = true]{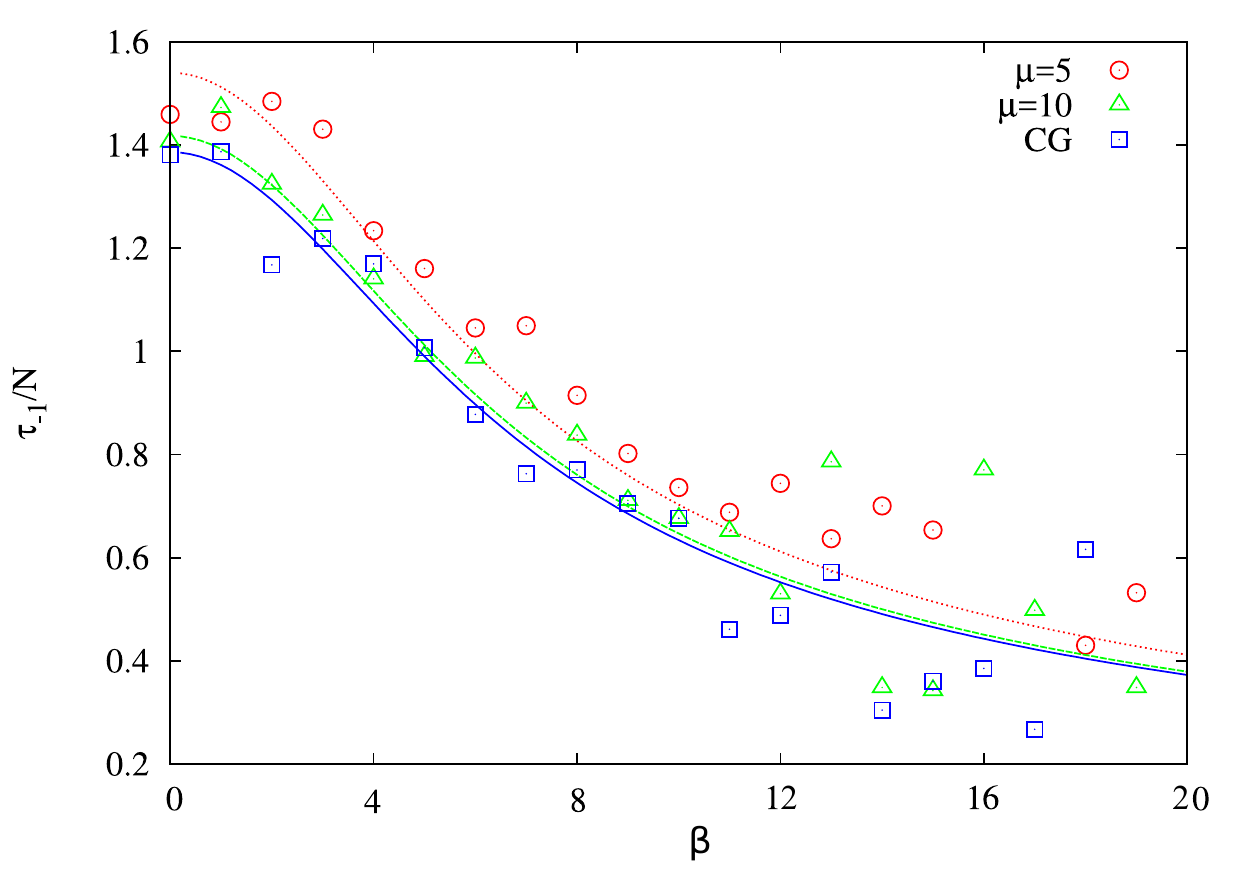}
\label{er_cg_taum1}
}
\caption{Average times $\tau_{\pm1}(\sigma)$ to reach consensus in $+1$ or $-1$ states, in units of MCS and rescaled by the system size $N$, for an initial condition $\sigma=0.5$ as a function of $\beta=\gamma v N$. Numerical results (symbols) are compared with analytical predictions (solid lines) both for the complete-graph (CG) and for the random Erd\H{o}s-R\'eny networks with different average degree $\mu$, as given by Eq.(\ref{ttrc1_8}) and Eq.(\ref{eq:tau1_network}) for $\tau_1$ and Eq.(\ref{ttrc_9}) for $\tau_{-1}$. Same line, symbol meanings and parameter values than in Fig.~\ref{er_cg_ttrc}.}
\label{er_cg_ttrc1}
\end{figure}

The average time to reach the non-preferred state, $\tau_{-1}(\sigma;\beta)$ can be found by noting the relation :
\begin{equation}\label{ttrc_9}
\tau(\sigma;\beta)=P_1(\sigma;\beta)\tau_1(\sigma;\beta)+(1-P_1(\sigma;\beta))\tau_{-1}(\sigma;\beta)
\end{equation}
and using Eqs.(\ref{p1_9}, \ref{ttrc_6}, \ref{ttrc1_8}). 
We show in Figs \ref{er_cg_tau1}, \ref{er_cg_taum1} that the analytical expression for the times to reach the absorbing states $\tau_{1}$ or $\tau_{-1}$ agree well with computer simulations of the system dynamics both in the complete graph and in the random networks to be discussed in section \ref{sec:random}. The presence of a bias decreases both times $\tau_{1}$ and $\tau_{-1}$. 

\section{Random networks}\label{sec:random}
So far, we have only discussed the situation of a complete graph. In this section we want to consider more general networks of interactions. A network is fully defined through its adjacency matrix ${\cal A}$ whose elements are ${\cal A}_{i,j}=1$ if nodes $i$ and $j$ are connected and ${\cal A}_{i,j}=0$ otherwise. This detailed information is most times simplified to the knowledge of the degree distribution $P_k=N_k/N$, being $N_k$ the number of nodes with degree $k$ and $N$ the total number of nodes. The average degree is $\mu=\sum_{k} k P_k$, and the second moment $\mu_2=\sum_{k} k^2 P_k$. A random or connected Erd\H{o}s-R\'eny network is constructed by linking each possible pair of nodes with a given probability $p$. In the large $N$ limit such a network follows a Poisson distribution for $P_k$, $k>0$, with an average value $\mu=p N$ and second moment $\mu_2=\mu^2+\mu$. As usual, in the numerical simulations we disregard those networks that can be split in two disconnected parts.

\subsection{Pair approximation}\label{Sec:pair_approx}
At the level of the mean-field approximation, the only relevant variable is the fraction $\sigma(t)$ of sites in the state $s=+1$ as a function of time. Within this mean-field approximation, the probability that a randomly selected pair of neighbors is ``active'', i.e. both sites are in different states, is $2\sigma(1-\sigma)$, a result coming from the statistical independence assumed in the approximation. At a more detailed level, the {\slshape pair approximation} considers correlations between the states of different connected sites by introducing the density of active links, $\rho(t)$, as a new dynamical variable~\cite{2008Vazquez,Pugliese:2009,2010Vazquez,2018bPeralta,Peralta2020b}. This approach is reasonably successful to treat random network configurations without degree-degree correlations such as an Erd\H{o}s-R\'eny network. If $i,j$ are connected nodes, we define $\rho^{i,j}=\text{Prob}(s_i=-s_j)$ as the probability that the link $i,j$ is active. The global density of active links is then $\rho=\frac{1}{L}\sum_{\langle i,j\rangle}\rho^{i,j}$, being $L=\mu N/2$ the total number of links. The pair approximation assumes that the probability of a link being active is independent of the state of the other links, hence $\rho^{i,j}\approx \rho$. Consistent with this approximation it is further assumed that $\text{Prob}(s_i=-s,s_j=s)\approx \rho/2$ independently of the value of $s=\pm1$.

Beyond the mean-field approach developed in the previous sections, the pair approximation uses $\rho(t)$ and $\sigma(t)$ as an independent pair of dynamical variables to describe the state of the system. Note, however, that $\rho(t)=0$ is only consistent with $\sigma(t)=0,1$, coming from the fact that a consensus state, one in which all nodes hold the same value of their state variable, has no active links. It is of course possible to include further variables in the analysis. For instance, the set of degree-dependent fractions defined as the ratio $\sigma_k=n_k/N_k$ between the number $n_k$ of nodes which are in state $+1$ and have degree $k$ and the total number $N_k$ of nodes with degree $k$. It is obviously $\sigma=\sum_k P_k\sigma_k$. A better description of the state of the network replaces the fraction of nodes in the state $+1$ by the {\slshape degree-weighted fraction} $\sigma_L=\frac{1}{\mu}\sum_k P_k k \sigma_k$. For a regular or all-to-all connected network where $P_k=\delta_{k,\mu}$, the degree-weighted fraction $\sigma_L(t)$ coincides with $\sigma(t)$. 

A complete and closed description of the evolution of the dynamical variables $\rho(t),\,\sigma(t),\,\sigma_L(t)$ is possible within the context of the pair approximation. This description is, however, rather cumbersome~\cite{2018Peralta} and we have decided to present here a simplified treatment based on~\cite{2008Vazquez}. The idea is to consider a random walk not for the variable $\sigma(t)$ but for the variable $\sigma_L(t)$. A microscopic update $s_i\leftarrow s_j$ where node $i$ with degree $k_i=k$ copies the state of node $j$ modifies $\sigma_L$ by an amount $\pm\dfrac{k}{\mu N}\equiv \pm \Delta_k$. We now compute the probability $R_k^+$ that, given that node $i$ with degree $k_i=k$ has been chosen for updating, the change of the degree-weighted fraction $\sigma_L$ is $+\Delta_k$:
\begin{eqnarray}
R_k^+&=&\text{Prob}(s_i=-1,s_j=1|k_i=k)\left[\text{Prob}(i \text{ no bias})\frac{1}{2}+\text{Prob}(i\text{ bias})\frac{1+v}{2}\right]\nonumber\\
&=&\frac{\rho}{2}\frac{1+\gamma v}{2},\label{Rnet}
\end{eqnarray}
where we have used the approximation $\text{Prob}(s_i=-1,s_j=+1|k_i=k)=\text{Prob}(s_i=-1,s_j=+1)=\dfrac{\rho}{2}$ and that the probability that a node is biased or unbiased is independent of its degree. Similarly for the probability $R_k^-$ that, given that node $i$ with degree $k_i=k$ has been chosen for updating, the change of the degree-weighted fraction $\sigma_L$ is $-\Delta_k$ we obtain:
\begin{eqnarray}
R_k^-&=&\frac{\rho}{2}\frac{1-\gamma v}{2}. \label{Lnet}
\end{eqnarray}
Note that we still can define the rates $R^+(\sigma),\,R^-(\sigma)$ that the fraction $\sigma$ decreases or increases, respectively, by an amount $\Delta_\sigma=1/N$. Within this context, they are equal to $R_k^+$ and $R_k^-$, respectively, as these are independent of $k$ due to the approximations considered.

To proceed, we need an equation for the time evolution of $\rho$. We follow closely the derivation of~\cite{2008Vazquez} and note that every time a node with degree $k$ and $\ell$ active links is updated, the density of active links varies in an amount $\Delta \rho=\dfrac{2(k-2\ell)}{\mu N}$. As time increases by $\Delta_t=1/N$ after every node update, we write:
\begin{eqnarray}
\frac{d\rho}{dt}&=&\sum_kP_k\left.\frac{\Delta \rho}{\Delta_t}\right|_k \\ \nonumber
&=&\sum_k\frac{P_k}{1/N}\sum_{s=\pm1} P(s\to-s)\sigma_s\sum_{\ell=0}^kB(\ell,k|s)\frac{\ell}{k}\frac{2(k-2\ell)}{\mu N},
\label{rhot1}
\end{eqnarray}
where $\left.\dfrac{\Delta\rho}{\Delta_t}\right|_k$ denotes the average change in $\rho$ when a node of degree $k$ is chosen, $P(s\to -s)$ is the probability that the proposed change $s\to-s$ is accepted, and $B(\ell,k|s)$ is the conditional probability that $\ell$ of the $k$ links connected to a node are active, given that the node is in the state $s$. We have introduced the notation $\sigma_1=\sigma$, $\sigma_{-1}=1-\sigma$. This expression is equivalent to
\begin{equation}
\frac{d\rho}{dt}=\frac{2}{\mu}\sum_k P_k \sum_{s=\pm 1}P(s\rightarrow -s)\sigma_{s}\left(\langle \ell\rangle_{k,s}-\frac{2}{k}\langle \ell^2\rangle_{k,s} \right)
\label{rhot}
\end{equation}
where $\langle \ell\rangle_{k,s}$, is the average number of active neighbors of a node in state $s$ and degree $k$. Using the pair approximation and neglecting correlation of second and higher neighbors, it turns out that $B(\ell,k|s)$ becomes a binomial distribution, whose first and second moments are 
\begin{eqnarray}
\langle \ell\rangle_{k,s} =\frac{\rho k}{2\sigma_{s}},\,
\langle \ell^2\rangle_{k,s} =\frac{\rho k}{2\sigma_{s}}+\frac{k(k-1)\rho^2}{4\sigma_{s}^2}.
\end{eqnarray}
Replacing in Eq.(\ref{rhot}), using $P(-1\rightarrow 1)=\frac{1+\gamma v}{2}$, $P(1\rightarrow -1)=\frac{1-\gamma v}{2}$, we arrive at
\begin{equation}
\frac{d\rho}{dt}=\frac{2\rho}{\mu}\left[\left(\frac{\mu}{2}-1\right)-\rho\left(\mu-1\right)\frac{1+\gamma v(2\sigma -1)}{4\sigma\left(1-\sigma\right)}\right].
\label{eq:rho}
\end{equation}
This equation has to be combined with the evolution equation for the fraction $\sigma$:
\begin{equation}
\frac{d\sigma}{dt}=R^+(\sigma)-R^-(\sigma)=\frac{\gamma v}{2}\rho.
\label{eq:sigma}
\end{equation}

The set of coupled equations (\ref{eq:rho}) and (\ref{eq:sigma}) are the basis of our subsequent analysis. They are the result of the pair approximation which neglects finite size fluctuations and it is therefore valid in the thermodynamic limit. In this limit and in the absence of bias ($v=0$), $\sigma$ is a conserved quantity and there is a stationary solution with a finite value of $\rho$. However, when bias is present, the stationary solution fulfills $\rho=0$, indicating that the absorbing state is reached by the intrinsic dynamics of the system in the absence of finite size fluctuations.
The dynamical equations (\ref{eq:rho}) and (\ref{eq:sigma}) reproduce the ones obtained in the preferred language study~\cite{2010Vazquez}, when setting $\gamma=1$, i.e. when all agents are biased, although our analysis is different. Instead of finding the general solution $\sigma(t),\,\rho(t)$ with given boundary conditions, we note that for $\gamma v$ small the time scale of Eq.(\ref{eq:sigma}) indicates that $\sigma(t)$ is a slow variable, and we assume that the dependence of $\rho(t)$ in time is through the relation $\rho(t)=\rho(\sigma(t))$. Dividing Eq.(\ref{eq:rho}) by Eq.(\ref{eq:sigma}) we get a closed differential equation to find the dependence $\rho(\sigma)$:
\begin{equation}
\frac{d\rho}{d\sigma}=\frac{4}{\mu \gamma v}\left[\left(\frac{\mu}{2}-1\right)-\rho\left(\mu-1\right)\frac{1+\gamma v(2\sigma -1)}{4\sigma\left(1-\sigma\right)}\right].
\label{eq:rho_sigma}
\end{equation}
The solution satisfying the boundary conditions $\rho(\sigma=1)=\rho(\sigma=0)=0$ is
\begin{eqnarray}
\rho(\sigma)&=&\frac{\mu-2}{\mu-1+\gamma v}2\sigma(1-\sigma)^{a_1}{_2F_1}\left(a_1,a_2;1+a_2;\sigma\right),
\label{eq:sol_sigma}\\
a_1&=&\frac{1+\gamma v}{\gamma v}\frac{\mu-1}{\mu},\nonumber\\
a_2&=&\frac{\gamma v +\mu-1}{\gamma v \mu},\nonumber
\end{eqnarray}
where ${_2F_1}(\cdot)$ is the hypergeometric function.
It is possible to check the limits
\begin{eqnarray}
\lim_{v\to0}\rho(\sigma)&=&\frac{\mu-2}{\mu-1}2\sigma(1-\sigma),\quad\forall \mu,\\
\lim_{\mu\to\infty}\rho(\sigma)&=&2\sigma(1-\sigma), \quad\forall \gamma v.
\end{eqnarray}
While the first limit coincides with the one obtained in~\cite{2008Vazquez}, the last limit is an important check of the consistency of the calculation. When $\mu= N-1$ every two nodes are connected and it follows the exact relation $\rho=\dfrac{2\sigma(1-\sigma)}{1-1/N}$ and, for $N\to\infty$, $\rho=2\sigma(1-\sigma)$ is mandatory independently on the value of the bias parameter $v$ or the fraction of biased agents $\gamma$.

We simplify the complicated functional relation $\rho(\sigma)$ as given in Eq.(\ref{eq:sol_sigma}) in order to use it in further calculations and get full analytical expressions for $P_1$ and $\tau$. To this end we use the previous asymptotic limits and expand around $\sigma=1/2$:
\begin{equation}
\frac{4\sigma(1-\sigma)\rho(1/2)}{\rho(\sigma)}=1-c_1(\mu,v)(2\sigma-1)+O((2\sigma-1)^2).
\end{equation}
While $\rho(1/2)$ and $c_1(\mu,v)$ can be fully expressed in terms of the hypergeometric function, it is possible to use approximate expressions valid for small $v$, namely $\rho(1/2)\approx \dfrac{\mu-2}{2(\mu-1)}$ and $c_1\approx\dfrac{\gamma v}{\mu-1}$. This leads to the approximation
\begin{equation}\label{rho_sigma_ap}
\rho(\sigma)\approx \frac{2(\mu-2)\sigma(1-\sigma)}{\mu-1+\gamma v(1-2\sigma)}.
\end{equation}

\begin{figure}[ht]
\subfigure[$\rho(\sigma)$]{
\includegraphics[width = 0.4\columnwidth, valign=c]{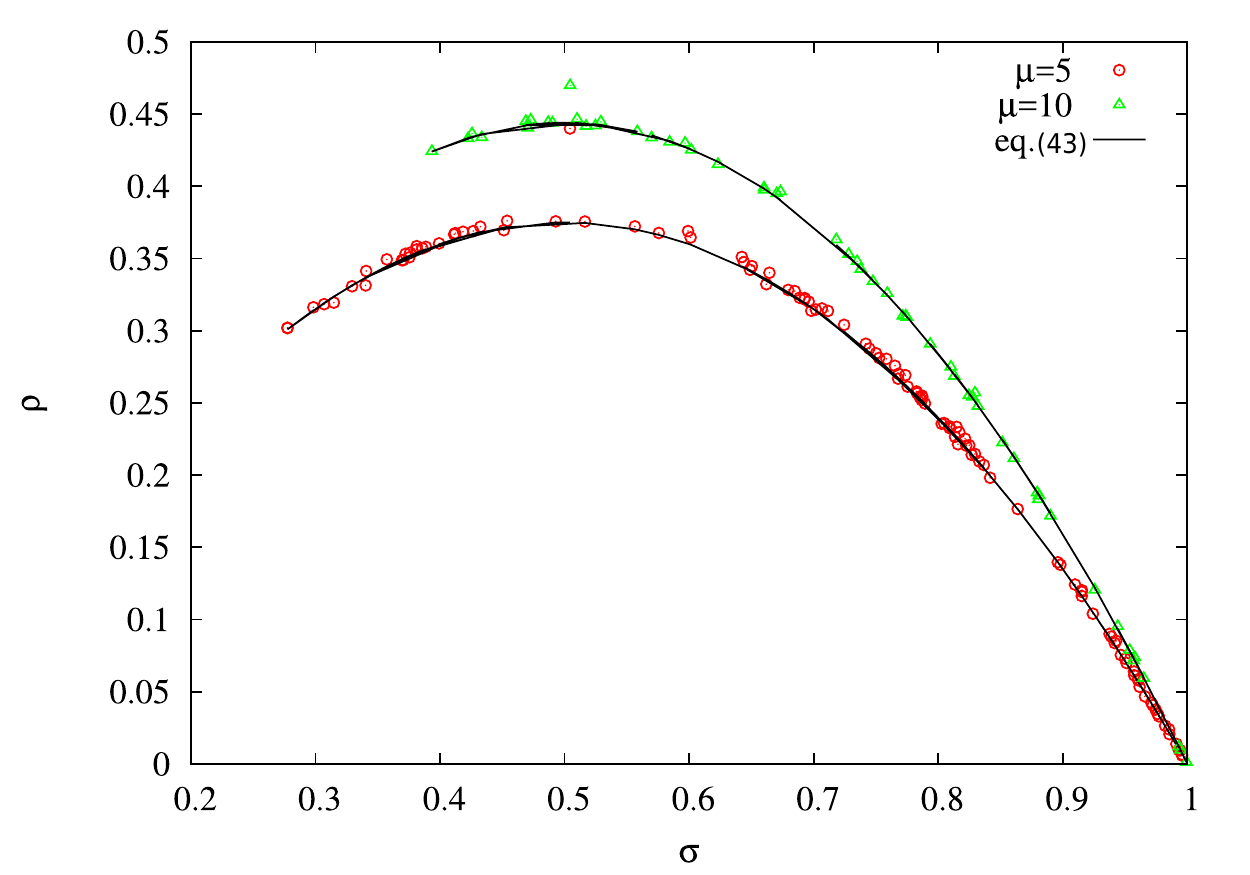}
\label{fig_er_rho}
}
\subfigure[$\rho(t)$]{
\includegraphics[width = 0.6\columnwidth, valign=c]{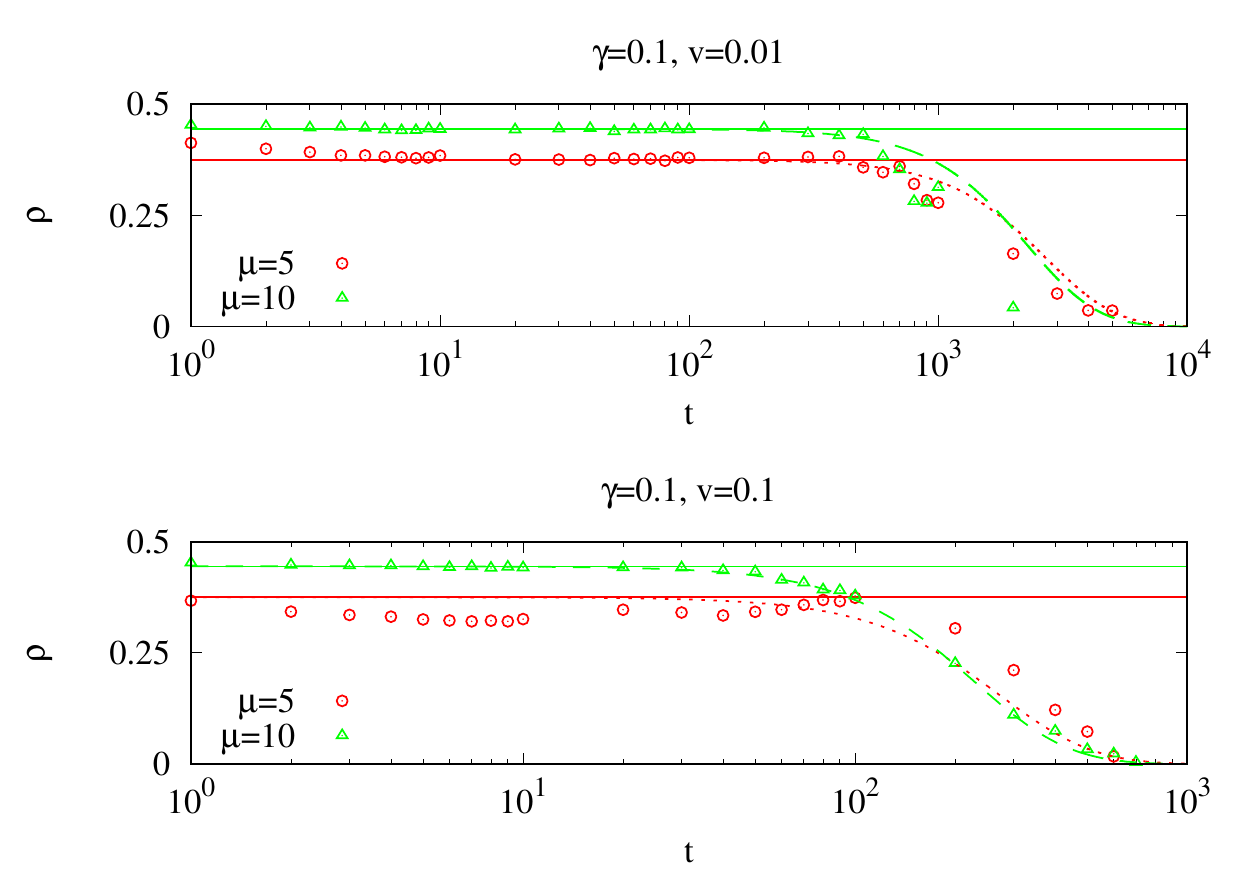}
\label{fig_er_time}
}
\caption{Density of active links $\rho$ for a single realization of the system dynamics of $N=10000$ agents on an Erd\H{o}s-R\'eny network and with initial conditions $\sigma(0)=1/2$. In panel $(a)$ we display with symbols $\rho(\sigma)$ as obtained from a parametric plot of the numerical simulation values $(\sigma(t),\rho(t))$ for average connectivity $\mu=5$ (circles) and $\mu=10$ (triangles) . The solid lines correspond to the adiabatic approximation Eq.(\ref{eq:sol_sigma}). Panel $(b)$ shows with symbols the time evolution of $\rho(t)$ coming from the numerical simulations. The corresponding solid lines are the result of the numerical integration of Eqs.(\ref{eq:rho},\ref{eq:sigma}). The horizontal solid lines correspond to the plateau values $\rho\approx \dfrac{\mu-2}{2(\mu-1)}$ as given by Eq.(\ref{rho_sigma_ap}) for $\sigma=1/2$. }
\end{figure}

The essence of our approximation, in contrast to other approaches in the literature~\citep{2008Vazquez,2010Vazquez}, is that $\rho(t)$ follows adiabatically $\sigma(t)$ and we can use Eq.(\ref{rho_sigma_ap}) using the time-dependent values $\sigma(t)$ and $\rho(t)$. The comparison with computer simulations shown in Fig. \ref{fig_er_rho} proves the goodness of this approximation for Erd\H{o}s-R\'eny networks and two different values of the average degree $\mu$. In Fig.~\ref{fig_er_time} we plot the time dependence of the interface density in a single realization of the dynamics of the system and compare it against the value of the plateau $\rho(1/2)\approx \dfrac{\mu-2}{2(\mu-1)}$ that follows from the stationary solution of Eq. (\ref{eq:rho})~\cite{2008Vazquez}. We now introduce this approximation to analyze the behavior of the fixation probability and the times to reach the different consensus states.

\subsection{Absorbing state}

In this subsection we go beyond the pair approximation, introducing finite size effects which are neglected in the previous treatment. In order to identify the absorbing state of the system, we resort to the relevant master equation for the probability $P(m, t)$ that the system has magnetization $m$ at time $t$. In a time step, a
node with opinion $s$ flips with probability $\sigma_sP(-s|s)P(s\rightarrow-s)$ after which
the magnetization $m=2\sigma-1$ changes by $\Delta_m=2s \Delta_k$ where $\Delta_k=\frac{k}{\mu N}$ with $k$ the number of neighbors of the selected node. 

Following a similar approach as before, we can arrive to the following Fokker-Planck equation 

\begin{eqnarray}
\frac{\partial P\left(m,t\right)}{\partial t}&=& 
-\frac{\mu-2}{\mu-1}\frac{\gamma v}{2}\frac{\partial \left[\frac{1-m^{2}}{1-\frac{\gamma v}{\mu-1}m}P\left(m,t\right)\right]}{\partial m}\\ &+&\frac{\mu-2}{\mu-1}\frac{1}{2N_\mu} 
\frac{\partial^{2}\left[\frac{1-m^{2}}{1-\frac{\gamma v}{\mu-1}m}P\left(m,t\right)\right]}{\partial m^{2}}\nonumber
\end{eqnarray}
where $N_{\mu}=\dfrac{\mu^2}{\mu_2}N$ or $N_{\mu}=\dfrac{\mu}{\mu+1}N$ for the Erd\H{o}s-R\'eny network. Here we see that if $\gamma$ or $v$ are set to 0 we obtain the same result as in~\cite{2008Vazquez}. We can now perform a similar analysis to that of Eq.(\ref{Fokker-Planck}) to conclude that for $\gamma v>0$ it is $P_\text{st}=\delta(m-1)$, and that the approach to the stationary state $m=1$ occurs in a characteristic time scale $\frac{\mu-1}{(\mu-2)\gamma v}$. 

\subsection{Fixation probability $P_1$}\label{sub:P1_networks}

To compute the fixation probability $P_1$ we consider the random walk in the degree-weighted fraction $\sigma_L$ which takes the value $\sigma_L=1$ in the consensus state $s_i=s,\,\forall i$. It can be computed by a reasoning similar to the one that led to Eq.(\ref{p1_1}), but considering the different contributions according to the degree of the selected node for updating:
\begin{eqnarray}
P_1(\sigma_L)=\sum_k P_k \biggl[ R_k^+P_1(\sigma_L+\Delta_k)+R_k^-P_1(\sigma_L-\Delta_k) +[1-R_k^+-R_k^-]P_1(\sigma_L)\biggr].
\label{p1_2}
\end{eqnarray}
Expanding up to second order in $\Delta_k=\dfrac{k}{\mu N}$ and replacing the rates (\ref{Rnet},\ref{Lnet}) we arrive at:
\begin{equation}
\frac{d^2 P_1}{d\sigma_L^2}+\beta_\mu \frac{dP_1}{d\sigma_L}=0,
\label{p1_8b}
\end{equation}
here $\beta_{\mu}=2\gamma v N_{\mu}$. Hence the solution Eq.(\ref{p1_9}) is still valid if we replace $\beta$ by $\beta_\mu$ and $\sigma$ by $\sigma_L$. Nevertheless, as proven by a more detailed analysis~\cite{2018bPeralta}, the variable $\sigma(t)$ follows $\sigma_L(t)$ and we can replace one variable by the other. However, it is essential to do the random-walk analysis in terms of the variable $\sigma_L$, otherwise the dependence on $N_{\mu}$ is lost. As shown in Fig.~\ref{fig_p1_cg_er}, where we plot the fixation probability for the complete-graph and two Erd\H{o}s-R\'eny networks with main degree $\mu=5$ and $\mu=10$, the agreement of Eq.(\ref{p1_9}) with the numerical results is very good if we include the system-size dependence in $N_{\mu}=\dfrac{\mu}{\mu+1}N$.

\subsection{Time to reach consensus, $\tau$} \label{sec_tau_er}

We modify the approach used for the complete-graph by noticing that in the case of a heterogeneous network, the change in the fraction of nodes in state $+1$ depends now on the degree $k$ of the node selected for update as $\Delta_k$. Hence we add all contributions weighted by its probability and modify Eq.(\ref{ttrc_1}) as
\begin{eqnarray}
\tau(\sigma_L)=&\sum_kP_k \biggl[R_k^+\left[\tau(\sigma_L+\Delta_k)+\Delta_t\right]+ R_k^-\left[\tau(\sigma_L-\Delta_k)+\Delta_t\right] \nonumber\\
&+\left[1-R_k^+-R_k^-\right]\left[\tau(\sigma_L)+\Delta_t\right]\biggr]\label{ttrc_1b}.
\end{eqnarray}
Expanding to second order in $\Delta_k=\frac{k}{\mu N}$ and replacing $\Delta_t=\frac{1}{N}$, this equation becomes 
\begin{equation}
\sum_{k} P_k\left[\frac{k^2}{2\mu^2N}\left(R_k^++R_k^-\right)\frac{d^2\tau}{d\sigma^2}+\frac{k}{\mu}\left(R_k^+-R_k^-\right)\frac{d\tau}{d\sigma} \right]=-1,
\label{ttrc_2b}
\end{equation}
and after replacing the rates as given by Eqs. (\ref{Rnet},\ref{Lnet},\ref{rho_sigma_ap}) 
\begin{equation}\label{eq:d2tauL}
\frac{d^2\tau}{d\sigma^2}+\beta_{\mu}\frac{d\tau}{d\sigma} =-2N_{\mu}\left(\frac{A}{\sigma}+\frac{B}{1-\sigma}\right),
\end{equation}
where
\begin{eqnarray}
A\equiv \frac{\mu-1+\gamma v}{\mu-2},\label{Aer} \\
B\equiv \frac{\mu-1-\gamma v}{\mu-2},\label{Ber}
\end{eqnarray} 
and, in view of the aforementioned equivalence, we have replaced $\sigma_L$ by $\sigma$. 
The solution of Eq.\eqref{eq:d2tauL} with the boundary conditions $\tau(0)=\tau(1)=0$ can be written in terms of the function $T(\sigma;\beta)$ defined in Eq.(\ref{ttrc_7c}) as 
\begin{equation}\label{eq:tau_network}
\tau(\sigma;\beta_{\mu})= 2N_{\mu}\left[AT(\sigma;\beta_{\mu})+ BT(1-\sigma;-\beta_{\mu})\right].
\end{equation}
In Fig. \ref{er_cg_ttrc} we compare this analytical solution with the results of computer simulations for Erd\H{o}s-R\'eny networks with average degree $\mu=5$, $\mu=10$. We observe that the time to reach consensus increases for decreasing average degree $\mu$ and that it is larger in an Erd\H{o}s-R\'eny network than in the all-to-all configuration. Note, again, that analyzing the random walk in terms of $\sigma$ instead of $\sigma_L$ we would have missed the dependence on $N_{\mu}$ which provides a much better fit to the numerical data.

\subsection{Time to reach preferred state, $\tau_1$}
We start by writing an analog expression of Eq.(\ref{ttrc1_1}) for heterogeneous networks (see Sec. \ref{sec_tau_er}) but considering a random walk in the $\sigma_L$ variable with contributions depending on the degree $k$
\begin{eqnarray}
P_1(\sigma_L)\tau_1(\sigma_L)&=&\sum_kP_k\left\{R_k^+\left[P_1(\sigma_L+\Delta_k)\tau_1(\sigma_L+\Delta_k)+P_1(\sigma_L)\Delta_t\right] \right.\nonumber
\\&&+ R_k^-\left[P_1(\sigma_L-\Delta_k)\tau_1(\sigma_L-\Delta_k)+P_1(\sigma_L)\Delta_t\right] 
 \nonumber\\ &&+\left.\left[1-R_k^+(\sigma_L)-R_k^-(\sigma_L)\right]\left[P_1(\sigma_L)\tau_1(\sigma_L)+P_1(\sigma_L)\Delta_t\right]\right\}.\label{ttrc1er_1}
\end{eqnarray}
Expanding to second order in $\Delta_k=\frac{k}{\mu N}$ and replacing $\Delta_t=\frac{1}{N}$, we obtain
\begin{eqnarray}
\sum_kP_k\biggl[\frac{k^2}{2\mu^2N}\left(R_k^++R_k^-\right)\frac{d^2(P_1\tau_1)}{d\sigma_L^2}
+\frac{k}{\mu}\left(R_k^+-R_k^-\right)\frac{d(P_1\tau_1)}{d\sigma_L}\biggr]=-P_1.
\label{ttrc1er_2}
\end{eqnarray}
Using in the right-hand-side the expression of $P_1(\sigma)$ from Eq.(\ref{p1_9}) with $\beta$ replaced by $\beta_\mu$, and substituting again $\sigma_L$ by $\sigma$, we obtain
\begin{eqnarray}
\frac{d^2(P_1\tau_1)}{d\sigma^2}+\beta_{\mu}\frac{d(P_1\tau_1)}{d\sigma}=-\frac{2N_{\mu}}{1-e^{-\beta_{\mu}}}\left(\frac{A}{\sigma}+\frac{B}{1-\sigma}-\frac{Ae^{-\beta_{\mu}\,\sigma}}{\sigma}-\frac{Be^{-\beta_{\mu}\,\sigma}}{1-\sigma}\right),
\label{tauer_general}
\end{eqnarray}
where $A$, $B$ are given by Eqs. (\ref{Aer},\ref{Ber}). The solution with boundary conditions $\tau_1(1)P_1(1)=\tau_1(0)P_1(0)=0$ can be expressed in terms of the function defined in Eq. (\ref{ttrc_7c}) as
\begin{eqnarray}\label{eq:tau1_network}
&&\tau_1(\sigma;\beta_{\mu})=\nonumber \\ &&\frac{2N_{\mu}}{1-e^{-\beta_{\mu}\,\sigma}}\biggl[AT(\sigma;\beta_{\mu})+BT(1-\sigma,-\beta_{\mu})
-Ae^{-\beta_{\mu}\,\sigma}T(\sigma;-\beta_{\mu})-Be^{-\beta_{\mu}\,\sigma}T(1-\sigma;\beta_{\mu})\biggr]\nonumber\\
\end{eqnarray}
In this case, it is no longer true that $\tau_1(\sigma=1/2;\beta_{\mu})=\tau(\sigma=1/2;\beta_{\mu})$, as it was in the mean-field approximation, although both times scale in the same way $\tau_1(\sigma=1/2;\beta_{\mu}),\tau(\sigma=1/2;\beta_{\mu})\sim \frac{B}{\gamma v}\ln(N_{\mu})$ as $N\to \infty$. In Fig.~\ref{er_cg_ttrc1} we compare these analytical results for $\tau_1(\sigma)$ and $\tau_{-1}(\sigma)$, derived from Eq.~(\ref{ttrc_9}) with the results of numerical simulations in Erd\H{o}s-R\'eny networks.

\section{Biased-dependent topology}

\label{Sec:biased}

So far we have assumed that links amongst agents, and hence the possibility of interaction, form, not just randomly, but also independently of whether they are biased or unbiased. In this section we will assume that the bias influences not only the interactions between individuals but, more importantly, the way they are connected, their network topology. Our goal is to determine whether the biased community is able to influence the whole system more efficiently by establishing their links in a more organized fashion. By more efficiently we mean that consensus to the preferred state occurs with a higher probability and in a shorter average time. To this end we will consider different network structures in which the connections between nodes will depend on their biased/unbiased label. 

For the sake of simplicity, we assume that the total number of links in the system, $L$, is fixed and given. Consequently, the total average degree of nodes $\mu=2L/N$ is also fixed. We denote the total number of links between biased-biased, unbiased-unbiased and biased-unbiased pairs of nodes as $\LBB$, $\LUU$, and $\LUB$ respectively. Thus the total number of links is $L=\LBB+\LUU+\LUB=\frac12 \mu N$. Let us denote by $\muB$, $\muU$ the average degrees of biased and unbiased nodes, respectively. They are related to the global average degree by $\mu=\gamma\muB+(1-\gamma)\muU$, if one assumes that the degree distribution $P_{k,U}$ for unbiased nodes is independent of the degree distribution $P_{k,B}$ for biased nodes. We can write the average degree of the given node as a sum of the connections to biased and unbiased neighbors, i.e., $\muB=\muBB+\muBU$ and $\muU=\muUB+\muUU$, where $\mu_{_\text{XY}}$ is the average number of links from an X-type node to a Y-type neighbor. Note that in general $\muUB$ and $\muBU$ are not equal but they are related as $\muBU \NB=\muUB \NU$ or $\gamma\muBU=(1-\gamma)\muUB$. For the generation of the biased-dependent topology networks we will use as a control parameter the ratio $\delta$ of links a biased node has to biased neighbors with respect to the number of links an unbiased node shares with its unbiased neighbors, i.e. $\delta=\dfrac{\muBB}{\muUU}$. In the case of a biased-independent random network this control parameter takes the value $\delta^\text{rand}=\dfrac{\gamma}{1-\gamma}$.

There are many ways in which one can modify the links in order to vary the parameter $\delta$ above or below its random value $\delta^\text{rand}$. Amongst all possibilities we have chosen to compare one case (so-called \textbf{model I}) in which the average connectivity of each agent remains always equal to $\mu$ and another case (so-called \textbf{model II}) in which the number of UB links is kept equal to that of the random network. In summary, in model I, we set $\muB=\muU=\mu$, while in model II we set $\muUB=\gamma \mu$. The fulfillment of these conditions, given the constraints listed in Table \ref{table_quantities}, leads, after a simple but lengthy algebra, to the values of the parameters listed in the same table.

\begin{table}[!h]
\begin{tabular}{| l l l l l ll |}
\hline \hline
quantity &\vline & Erd\H{o}s-R\'eny &\vline & model I &\vline & model II \\
\hline
$\muB$ &\vline& $\mu$& \vline& $\mu$ &\vline& $\dfrac{1+\delta -\gamma[2+\delta-\gamma(1+\delta)]}{1-\gamma(1-\delta)}\mu$ \\
\hline
$\muU$ &\vline& $\mu$ &\vline& $\mu$ &\vline& $\dfrac{1-\gamma[1-\gamma(1+\delta)]}{1-\gamma(1-\delta)}\mu$ \\ 
\hline 
$\muUU$ &\vline &$(1-\gamma)\mu$ &\vline& $\dfrac{1-2\gamma}{1-\gamma(1+\delta)}\mu$ &\vline& $\dfrac{(1-\gamma)^2+\gamma^2}{1-\gamma(1-\delta)}\mu$ \\
\hline
$\muUB$ &\vline& $\gamma\mu$ &\vline& $\dfrac{\gamma(1-\delta)}{1-\gamma(1+\delta)}\mu$ &\vline& $\gamma\mu$ \\
\hline
\hline
\end{tabular}

\begin{tabular}{|l|}

$\muBU=\dfrac{1-\gamma}{\gamma} \muUB$\\
\hline
$\muBB=\delta \muUU$\\
\hline
$\mu=\gamma\muB+(1-\gamma)\muB$\\
\hline
$\muB=\muBB+\muBU$\\
\hline
$\muU=\muUB+\muUU$\\
\hline
$\NB=\gamma N$ \\
\hline
$\NU=(1-\gamma)N$ \\
\hline
$\LBB=\frac12 \muBB N_B$ \\
\hline
$\LUU=\frac12\muUU N_U$ \\
\hline
$\LUB=\frac12( \muBU N_B+\muUB N_U)=\muBU N_B=\muUB N_U$ \\
\hline 
$L=\LUU+\LBB+\LUB=\frac12\mu N$ \\
\hline 
$\pUU=\muUU/N_U$ \\
\hline 
$\pBB=\muBB/N_B$ \\
\hline
$\pBU=\muBU/N_U=\muUB/N_B$ \\
\hline \hline
\end{tabular}
\caption{Comparison of relevant networks quantities, and definitions and relations among them, for an Erd\H{o}s-R\'eny network and the biased-dependent topologies. Setting the values of $\mu$, $\gamma$, $\delta$ and $N$ determines all other possible quantities. For the Erd\H{o}s-R\'eny network it is $\delta^\text{rand}=\frac{\gamma}{1-\gamma}$, while for model I and II, $\delta$ is a free parameter.}\label{table_quantities}
\end{table}

As, obviously, the quantities $\muUU,\,\muBB,\,\muUB,\,\muBU$ must be all non-negative, it follows from Table \ref{table_quantities} that when constructing model I it should be $\text{sign}(1-2\gamma)=\text{sign}(1-\delta)=\text{sign}(1-\gamma(1+\delta))$. A simple manipulation of these conditions allows us to conclude that there are two regions of allowed parameters $(\gamma$, $\delta)$ for model I, namely, $(\gamma<1/2,\delta<1)$, and $(\gamma>1/2,\delta>1)$. Although in the case of Model II we do not have any of those limitations, we restrict our posterior analysis to $\gamma<1/2$, where the biased community is a minority.

\begin{figure}[!h]
\centering
\hspace*{-0.2 in}
\subfigure[model I]{
\includegraphics[width = 0.5\columnwidth, keepaspectratio = true]{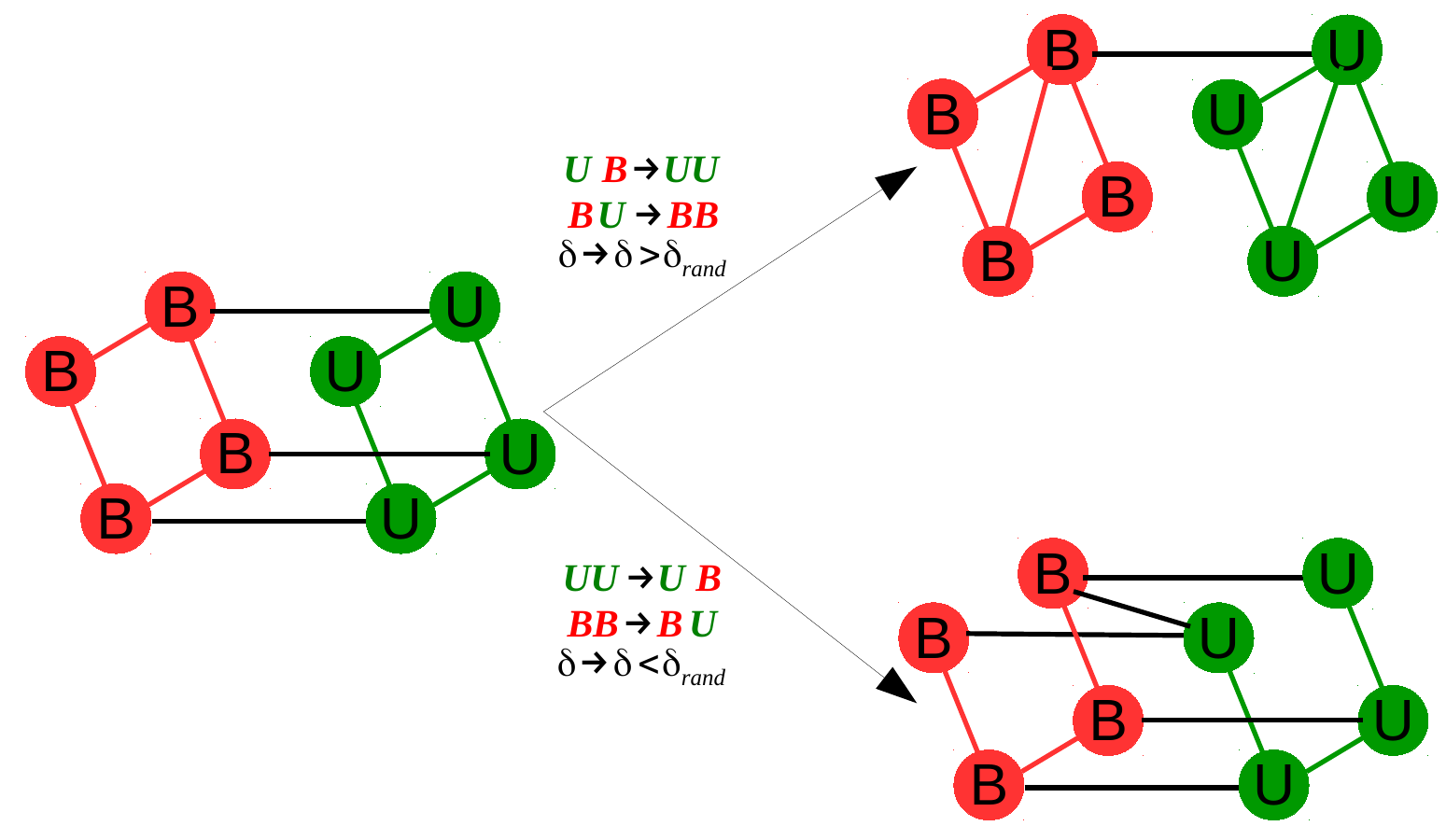}
\label{deltaI_strategy}
}
\hspace*{-0.2 in}
\subfigure[model II]{
\includegraphics[width = 0.5\columnwidth, keepaspectratio = true]{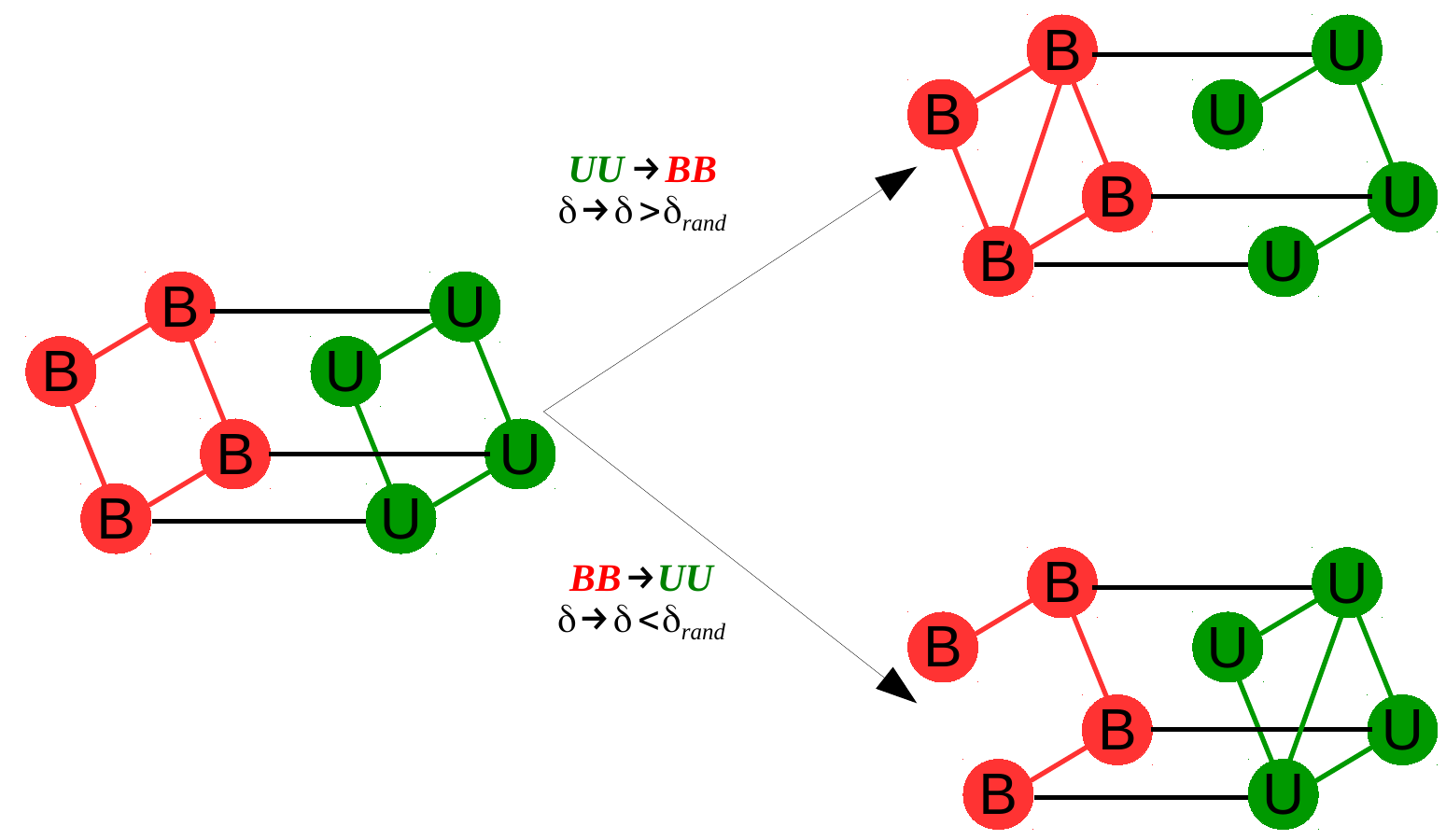}
\label{deltaII_strategy}
}
\caption{Schematic representation of the strategy to form biased-dependent topologies. Starting from a bias-independent Erd\H{o}s-R\'eny network, model I moves UB links to either UU or BB in order to increase $\delta>\delta^\text{rand}$, while the opposite moves of UU or BB to UB links decrease $\delta<\delta^\text{rand}$. This keeps the average connectivity of each agent unchanged. In model II the movement of a UU link to a BB type increases $\delta>\delta^\text{rand}$, while the opposite move of BB to UU decreases $\delta<\delta^\text{rand}$. This leaves constant the number of UB links joining both communities.}
\label{delta_strategy}
\end{figure}

\begin{figure}[!h]
\centering
\hspace*{-0.2 in}
\subfigure[$\gamma=0.2$ ($\delta^\text{rand}=0.25$)]{
\includegraphics[width = 0.5\columnwidth, keepaspectratio = true]{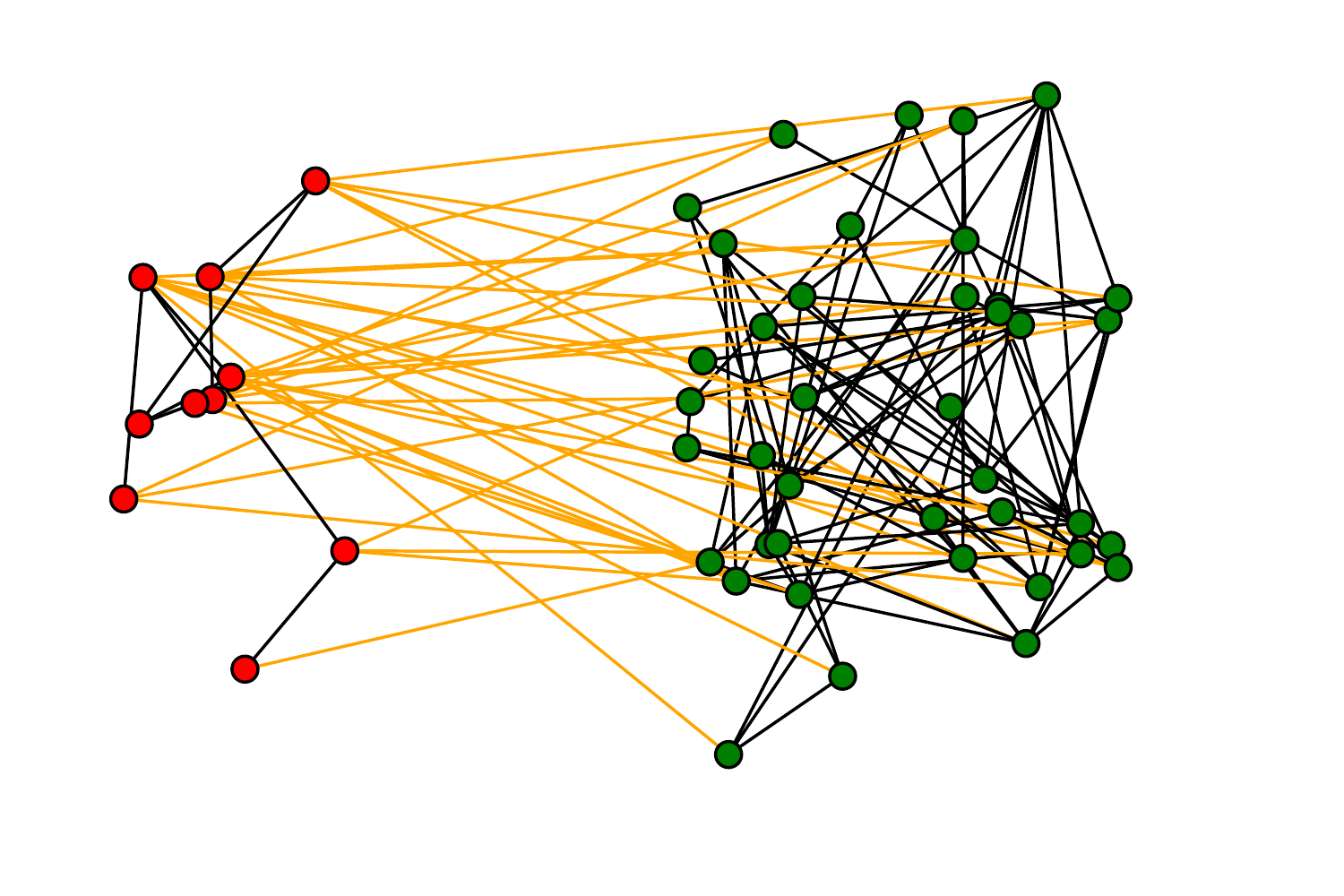}
\label{net_er_g02}
}
\hspace*{-0.2 in}
\subfigure[$\gamma=0.2$, $\delta=0.1<\delta^\text{rand}$]{
\includegraphics[width = 0.5\columnwidth, keepaspectratio = true]{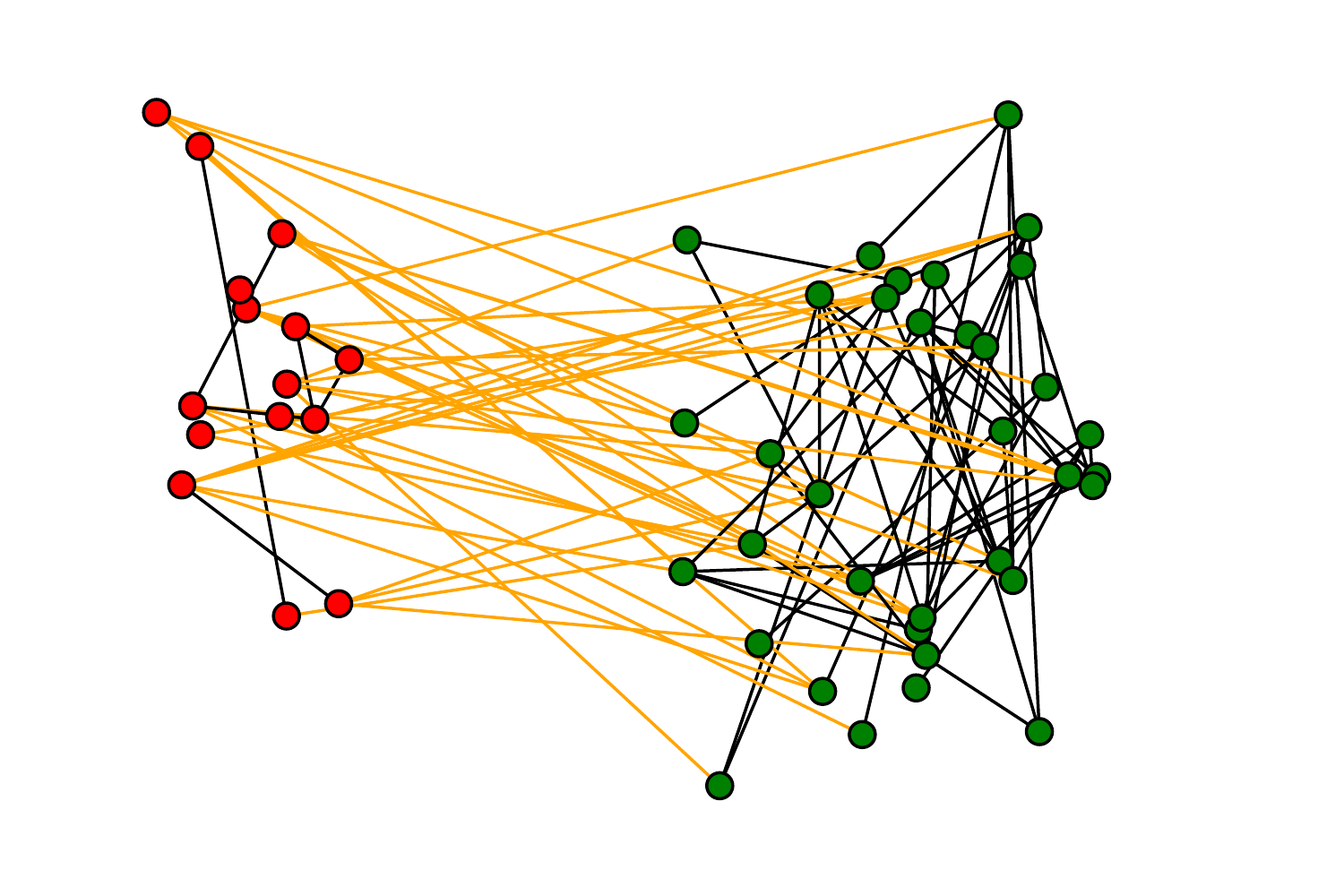}
\label{net_deltaII_m01g02}
}
\hspace*{-0.2 in}
\subfigure[$\gamma=0.2$, $\delta=10>\delta^\text{rand}$]{
\includegraphics[width = 0.5\columnwidth, keepaspectratio = true]{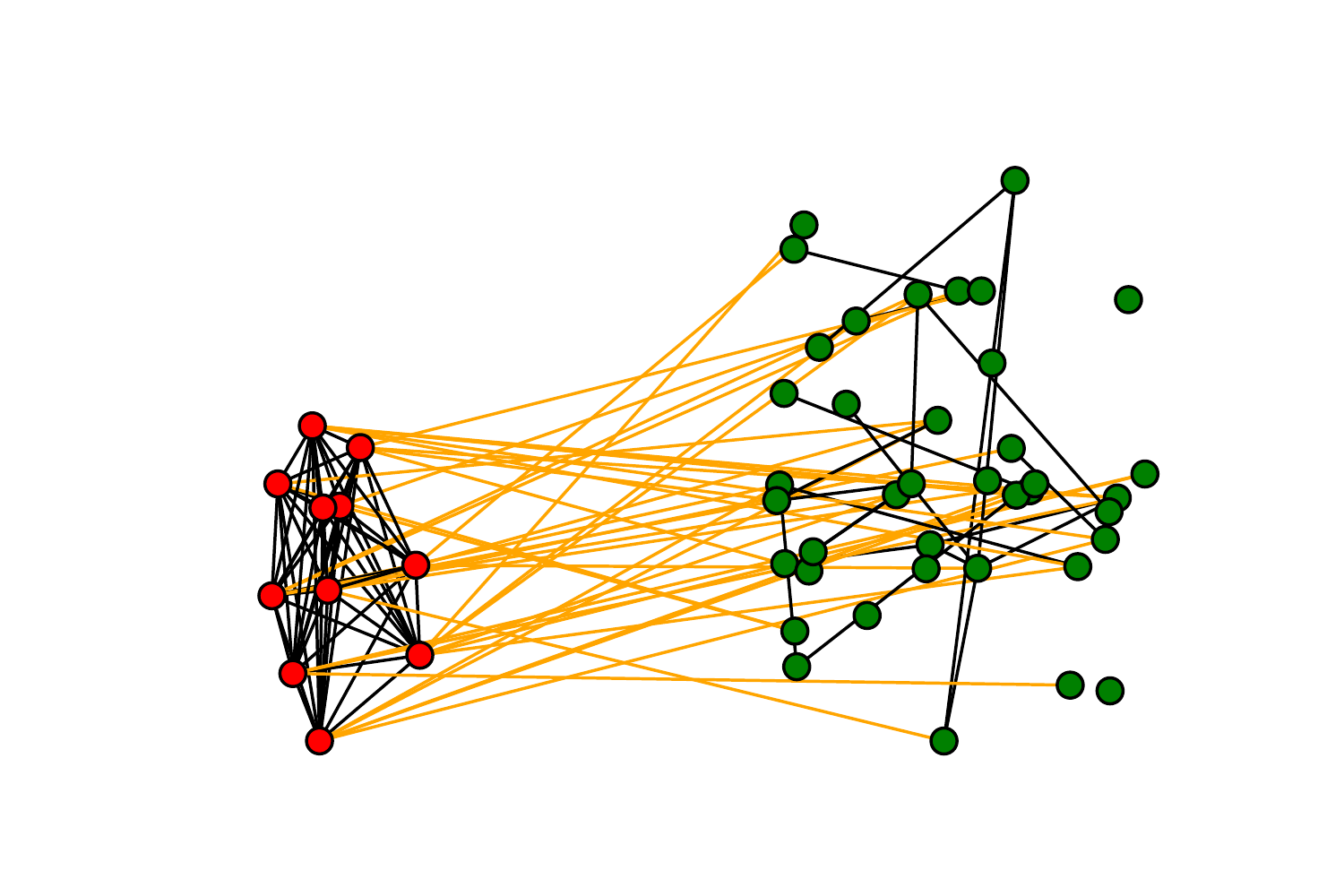}
\label{net_deltaII_m10g02}
}
\caption{Here we plot for Model II representative Erd\H{o}s-R\'eny networks of $N=50$ nodes with average degree $\mu=5$ with a $\gamma$ fraction of biased nodes. $\delta^\text{rand}=\frac{\gamma}{1-\gamma}$. On the left (red circles) we draw biased nodes and on the right (green circles) unbiased nodes. Black links connect nodes of the same type (i.e. UU and BB pairs), whereas orange links represent connections between pairs of nodes of different types (i.e. BU). }
\label{net_er}
\end{figure}

When analyzing the behavior of the average number of links $\muBB,\muUU,\muBU,\muUB$ as given in Table.~\ref{table_quantities}, for model I it turns out that if $\delta$ increases above the value $\delta^\text{rand}=\frac{\gamma}{1-\gamma}$, then $\muBB$ and $\muUU$ increase and $\muUB$ and $\muBU$ decrease with respect to the respective values $\muBB^\text{rand}=\gamma\mu$, $\muUU^\text{rand}=(1-\gamma)\mu$, $\muBU^\text{rand}=\gamma\mu$, $\muUB^\text{rand}=(1-\gamma)\mu$ they adopt in a random Erd\H{o}s-R\'eny network with average degree $\mu$. The opposite behavior, namely $\muBB<\muBB^\text{rand}$, $\muUU<\muUU^\text{rand}$, $\muBU>\muBU^\text{rand}$, $\muUB>\muUB^\text{rand}$, occurs for $\delta<\delta^\text{rand}$. In the case of Model II, $\muUB$ and $\muBU$ do not vary with $\delta$, but $\muBB$ increases and $\muUU$ decreases with respect to their random values $\muBB^\text{rand}$, $\muUU^\text{rand}$ when $\delta>\delta^\text{rand}$, and the opposite behavior $\muBB<\muBB^\text{rand}$, $\muUU>\muUU^\text{rand}$ if $\delta<\delta^\text{rand}$. This allows us to interpret Model I as follows: In order to increase $\delta>\delta^\text{rand}$, start from a random Erd\H{o}s-R\'eny network and rewire the necessary number of UB links (with equal probability) either as BB or as UU links. Analogously, in order to decrease $\delta<\delta^\text{rand}$, rewire an equal number of BB and UU links as UB links. Similarly, we can interpret model II as follows: In order to increase $\delta>\delta^\text{rand}$, start from a random Erd\H{o}s-R\'eny network and move UU links to BB links. Analogously, in order to decrease $\delta<\delta^\text{rand}$, move BB links to UU links. See Fig.~\ref{delta_strategy} for a schematic representation of these two strategies. In Fig.~\ref{net_er} we show some characteristic networks for values of $\delta$ equal, smaller and larger than the $\delta^\text{rand}$ value of the biased-independent Erd\H{o}s-R\'eny case in the case of model II.

In practice, those biased-dependent networks are constructed starting from $N$ disconnected nodes, a fraction $\gamma$ of which are biased, and linking each possible pair of nodes with probabilities $\pBB$, $\pUU$,or $\pUB$ if both nodes are biased, both nodes are unbiased, or one node is biased and the other unbiased, respectively. The random Erd\H{o}s-R\'eny network uses the same probability $p=\mu/N$ for the three cases. In order to achieve the correct network characteristics as before one must use the values of $\pBB,\,\pUU,\,\pUB$ listed in Table \ref{table_quantities}.

\begin{figure}[!h]
\centering
\includegraphics[width = \columnwidth, keepaspectratio = true]{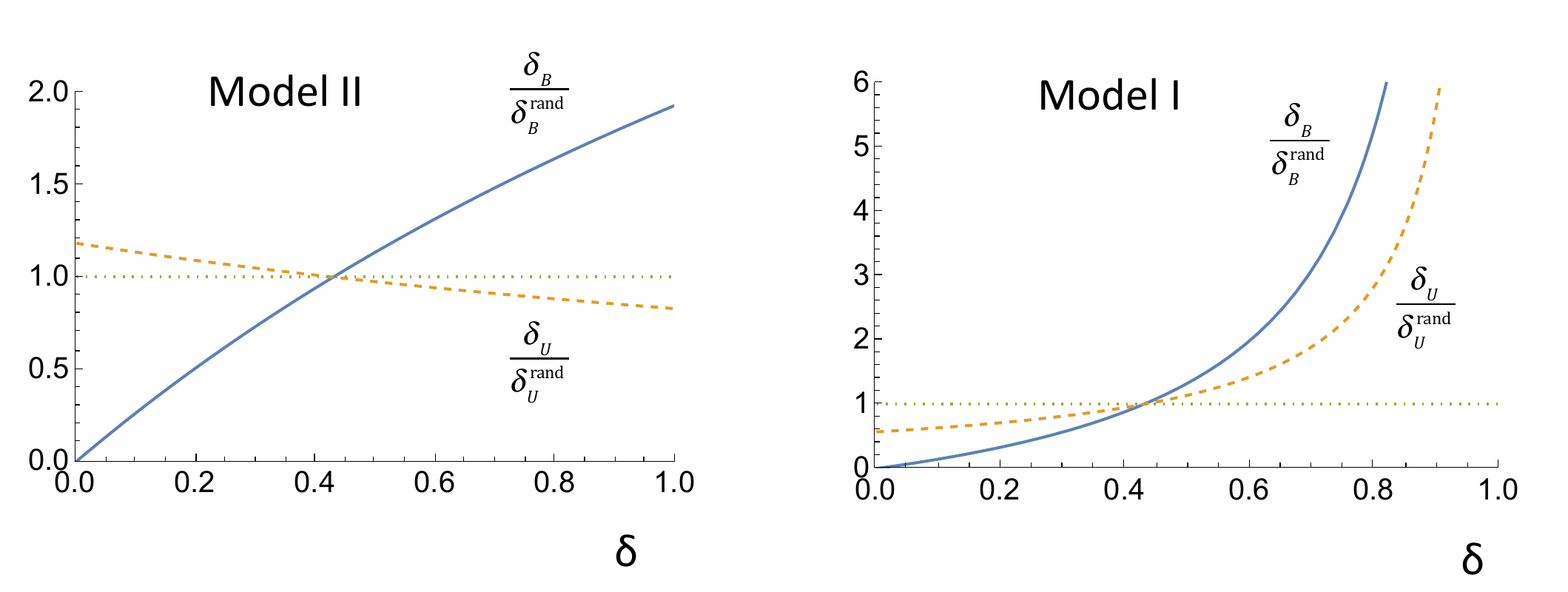}
\caption{$\deltaU\equiv\frac{\muUU}{\muUB}$ and $\deltaB\equiv\frac{\muBB}{\muBU}$, defined in terms of the average number of links between biased and unbiased communities, rescaled by the values they adopt in a random network where links are not determined by the preference of the agents, $\deltaB^\text{rand}=\frac{\gamma}{1-\gamma}$, $\deltaU^\text{rand}=\frac{1-\gamma}{\gamma}$, as a function of $\delta=\frac{\muBB}{\muUU}$ for both model I and II, defined in the main text. Values of $\deltaB$ larger (resp. smaller) than $\delta_B^\text{rand}$ indicate a closed (resp. open) biased community, and similarly for the unbiased community. We consider a minority fraction $\gamma=0.3$ of biased agents.}
\label{deltaUB}
\end{figure}

\begin{figure}[!h]
\centering
\hspace*{-0.2 in}
\subfigure{
\includegraphics[width = 0.5\columnwidth, keepaspectratio = true]{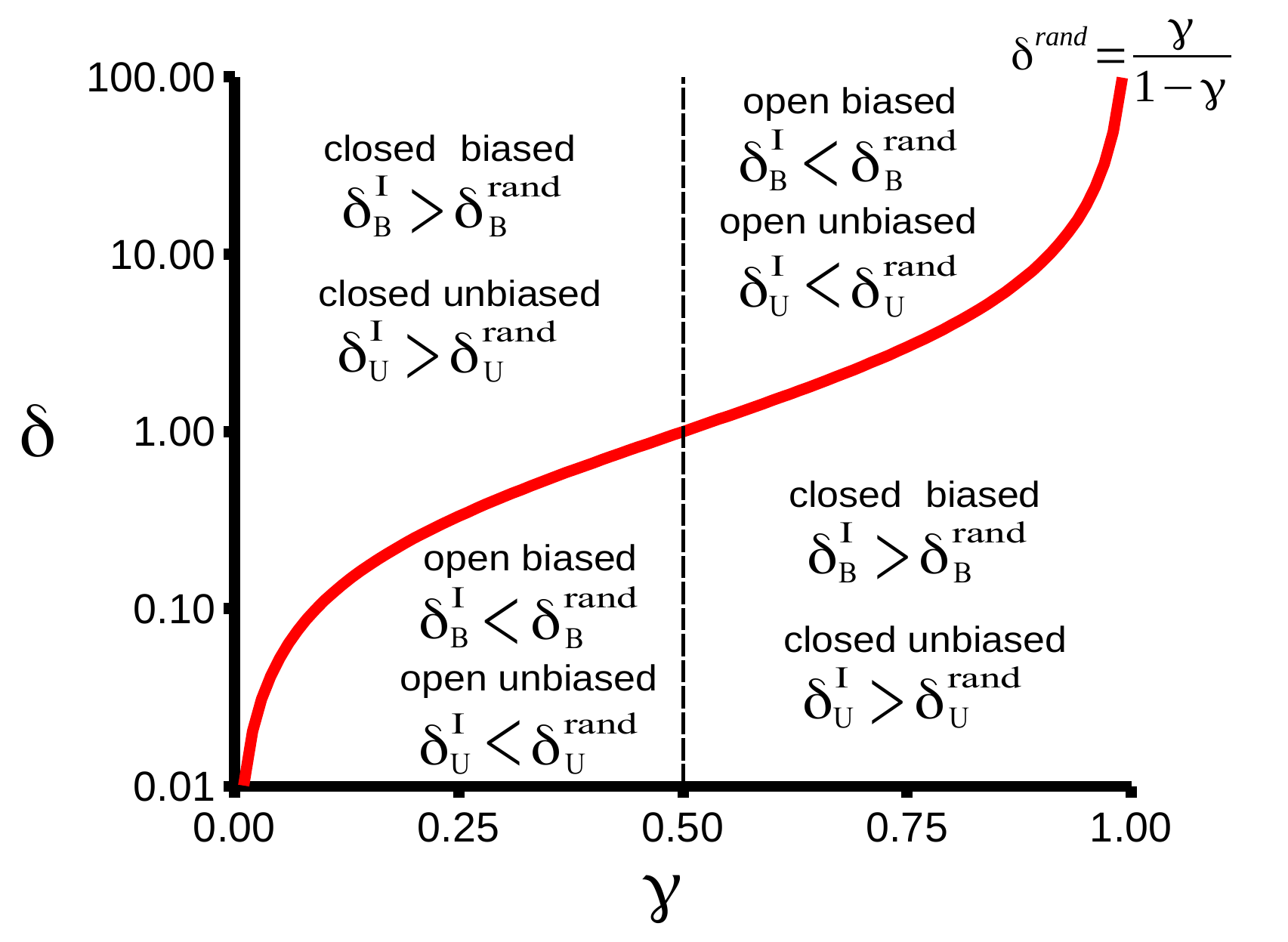}
\label{phase_diag_M1}
}
\hspace*{-0.2 in}
\subfigure{
\includegraphics[width = 0.5\columnwidth, keepaspectratio = true]{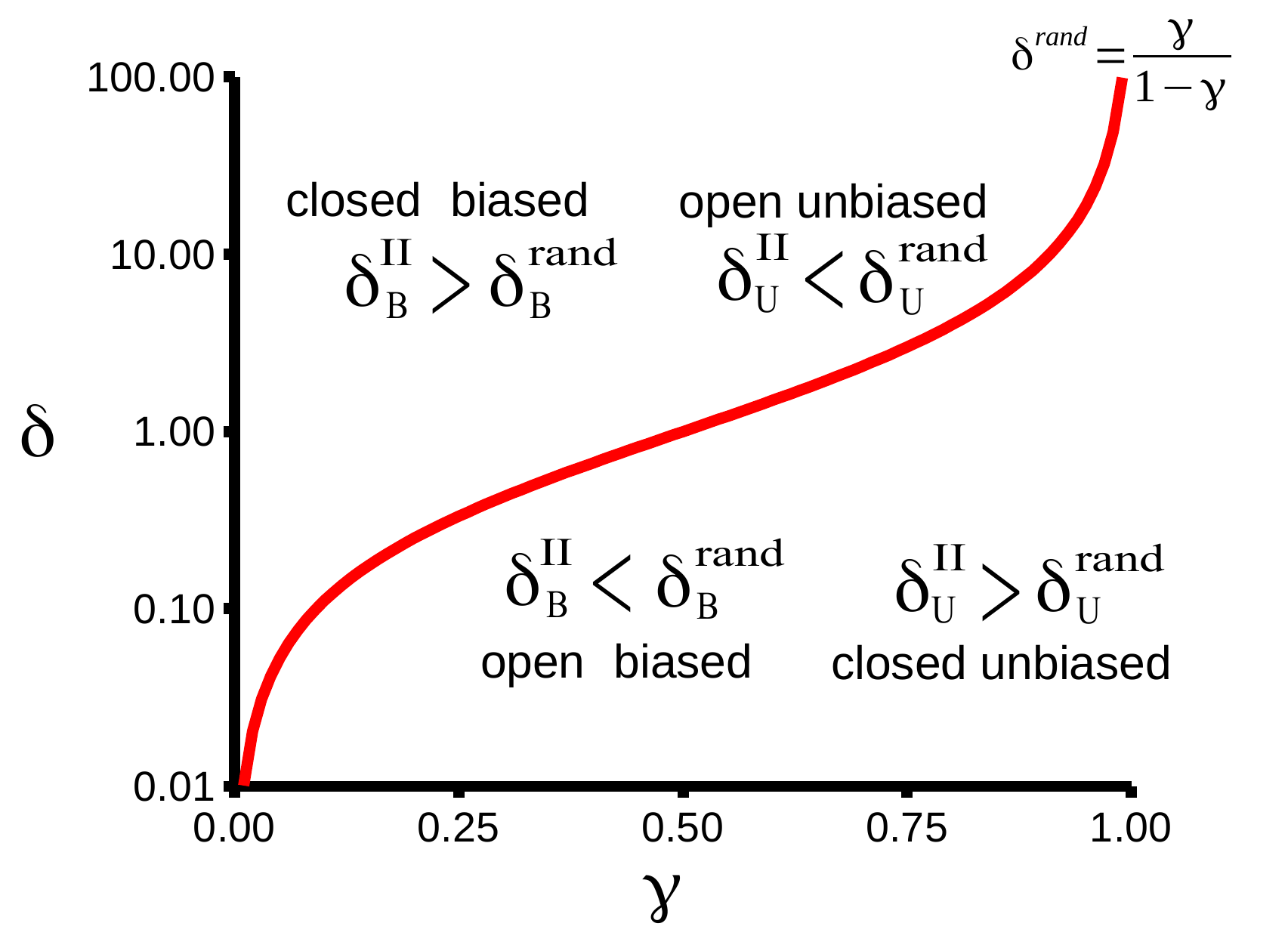}
\label{phase_diag_M2}
}
\caption{Here we plot the phase diagram according to the type, closed $\delta^\text{I,II}_\text{B,U}>\delta^\text{rand}_\text{B,U}$ or open $\delta^\text{I,II}_\text{B,U}<\delta^\text{rand}_\text{B,U}$, of each of the two communities, biased and unbiased, as a function of the parameters $\gamma$, the fraction of biased voters, and $\delta$, the ratio of the average number of connections among only biased voters over the average number of connections among only unbiased voters. On the left, we plot the phase diagram for model I. We see that in all of the phase space, the type of community is the same for both communities. We have a transition of the type when the line $\delta^\text{rand}$ is crossed, and also when the line $\gamma=0.5$ is crossed. On the contrary, for model II, the type of the two communities are always opposite, i.e. when the biased community is closed the unbiased is open and vice versa.}
\label{pd}
\end{figure}

We now discuss the type of communities that biased and unbiased agents form in each model for different values of the parameters $(\gamma,\delta)$. To this end we introduce, as a measure of how strongly united a community is, the ratio of the number of links that this community holds inside to the number of links it holds outside. For the biased community the measure is defined as $\deltaB\equiv 2\dfrac{\LBB}{\LUB}=\dfrac{\muBB}{\muBU}$, (the factor of $2$ in the definition is arbitrary) and for the unbiased community we use $\deltaU\equiv 2\dfrac{\LUU}{\LUB}=\dfrac{\muUU}{\muUB}$. For the biased-independent topology they adopt the values $\deltaB^\text{rand}=\frac{\gamma}{1-\gamma}$ and $\deltaU^\text{rand}=\frac{1-\gamma}{\gamma}$. Therefore, whenever $\deltaB>\deltaB^\text{rand}$, the biased community is more strongly linked internally than in the case that links are formed randomly without taking into account the preference of the agents, and we talk about a \textbf{closed} biased community. Similarly, for $\deltaB<\deltaB^\text{rand}$, the biased community has less internal links that those corresponding to a complete random assignment and we speak of an \textbf{open} biased community. A similar classification of \textsl{closed} or \textsl{open} applies to the community of unbiased agents for $\deltaU>\deltaU^\text{rand}$ or $\deltaU<\deltaU^\text{rand}$, respectively. As shown in Fig.~\ref{deltaUB}, it turns out that, for fixed $\gamma<1/2$, the biased community is closed for $\delta>\delta^\text{rand}$ and open for $\delta<\delta^\text{rand}$, independently of the model I or II considered. However, for model I, the unbiased community is open for $\delta<\delta^\text{rand}$ and closed for $\delta>\delta^\text{rand}$ and the opposite behavior for model II: open for $\delta>\delta^\text{rand}$ and closed for $\delta<\delta^\text{rand}$. This allows us to plot the phase diagram of Fig.~\ref{pd}, where, for completeness, we also include the characteristics of the communities for $\gamma>1/2$, a case not considered here.

Having established the strategies with which we obtain the biased-dependent topology of interactions of the network, we proceed to examine the dependence of the main observables, such as the fixation probability $P_1$, the consensus time $\tau$ and the time to reach consensus to the preferred state $\tau_1$, on the  different points of the phase diagram and its dependence with the closed/open property of the communities. As the results are rather representative, we focus on a fixed value of $\gamma=0.1$ and vary the parameter $\delta$ taking values smaller and larger than the purely random case $\delta^\text{rand}=\frac{\gamma}{1-\gamma}=\frac{0.1}{0.9}=0.11$. 

\begin{figure}[!h]
\subfigure[$\tau(\delta)$]{
\includegraphics[width = 0.50\columnwidth, keepaspectratio = true]{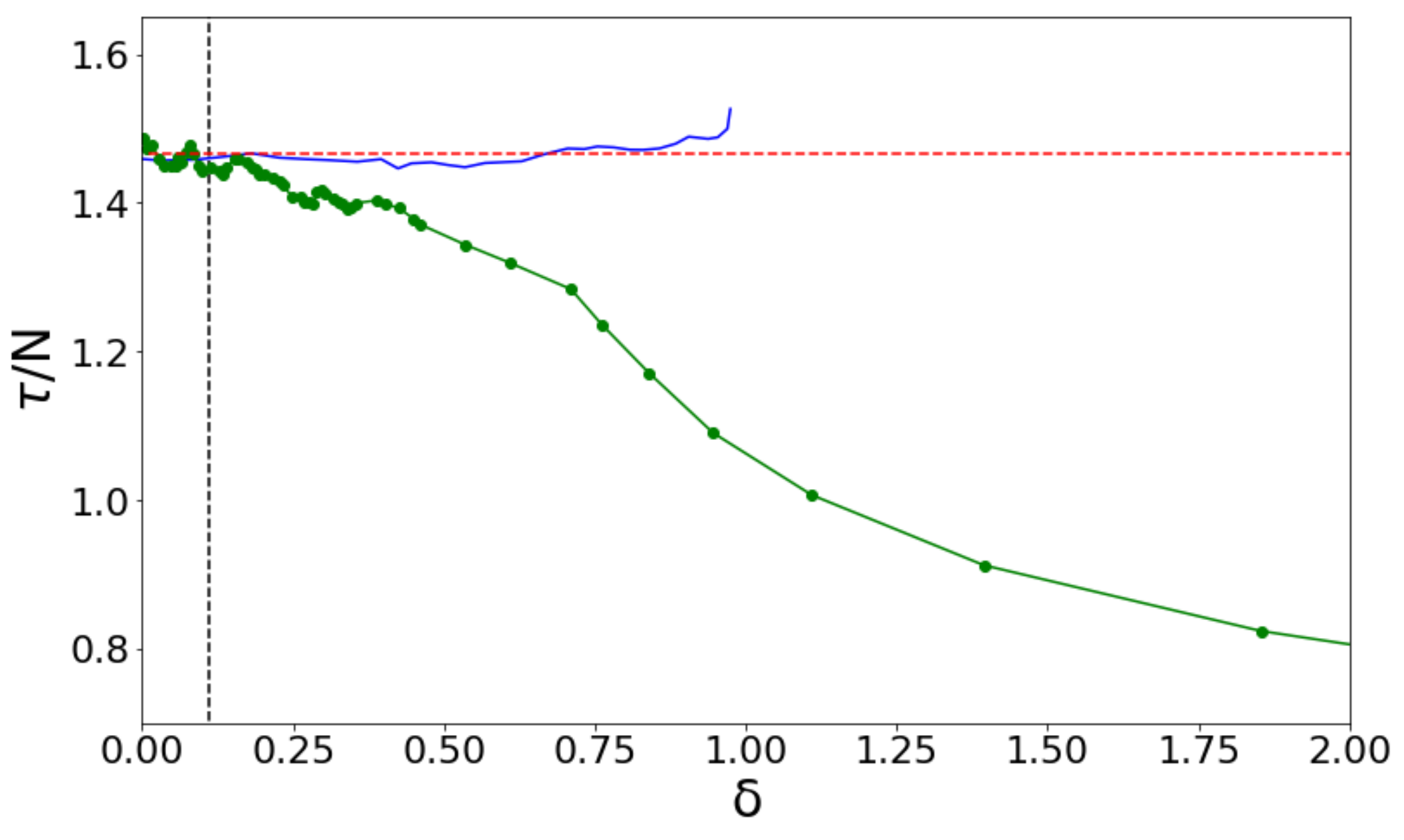}
\label{delta12_p1_ga}
}
\subfigure[$P_1(\delta)$]{
\includegraphics[width = 0.50\columnwidth, keepaspectratio = true]{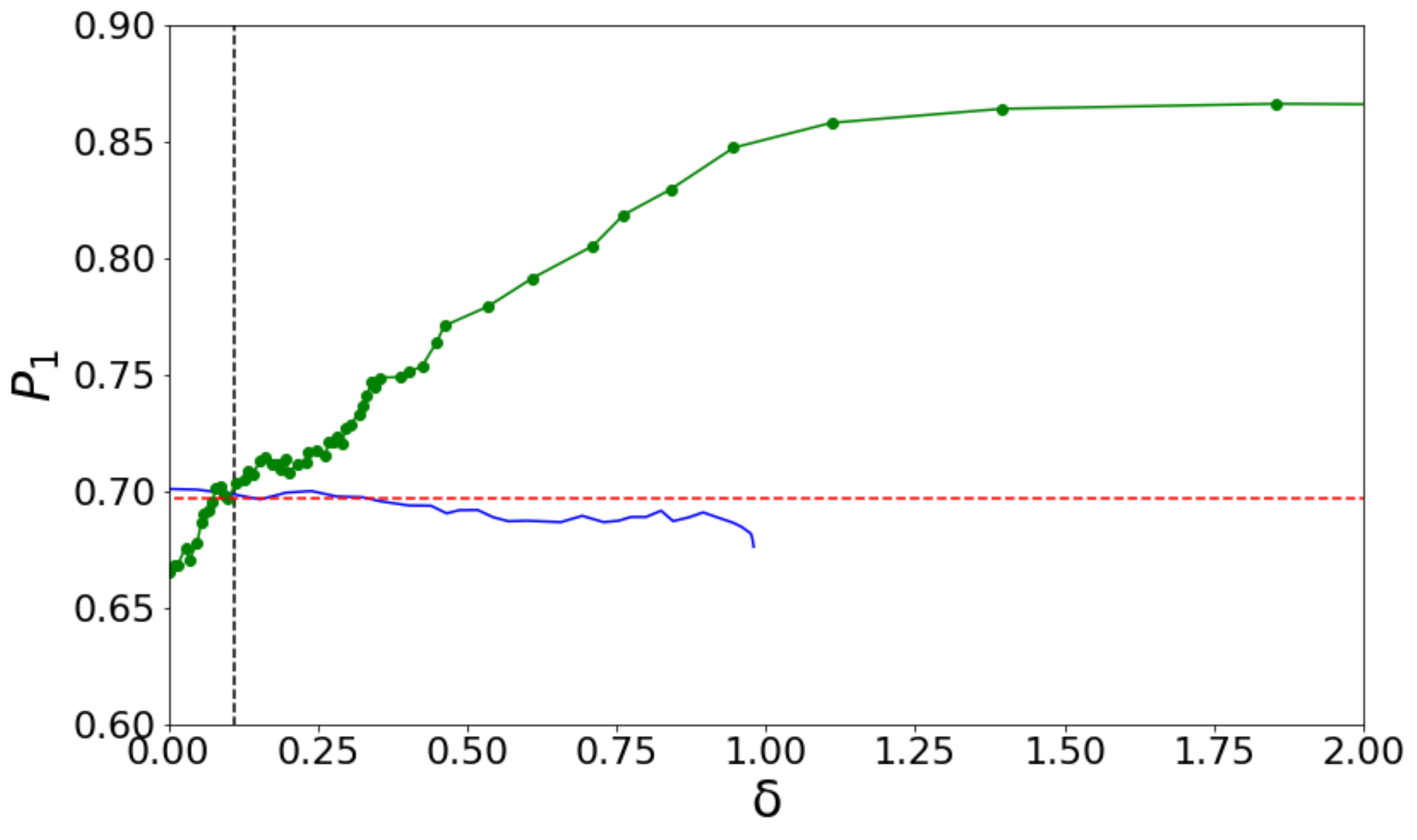}
\label{delta12_p1_gb}
}
\caption{(a) Time to reach consensus $\tau$ and (b) probability of reaching consensus in the preferred state $P_1$, as a function of the parameter $\delta=\frac{\muBB}{\muUU}$, starting from an initial condition $\sigma=0.5$. The red dashed horizontal lines corresponds to the theoretical result for an Erd\H{o}s-R\'eny network and a biased-independent topology, Eq.(\ref{eq:tau_network}) and Eq.(\ref{p1_9}), respectively, replacing $\beta$ with $\beta_\mu$, as explained in the main text. The vertical line is the value $\delta^\text{rand}=\frac{\gamma}{1-\gamma}$, that separates closed from open communities (see Fig.\ref{pd}). Blue continuous lines represent numerical results for model I (note that model I by construction is only defined up to $\delta=1$), while green dotted lines correspond to model II (see the text for a detailed explanation of these two models). We can see that model II results in a significant reduction of the consensus time as $\delta$ is increased, while it also results in a significant increase in the probability to reach consensus to the preferred state, when compared to both model I and the case of a homogeneous Erd\H{o}s-R\'enyi random network topology of interactions. The plot for the consensus time $\tau_1$ to the preferred state resembles (a) so it is omitted. The parameters are $v=0.01$, $\gamma=0.1$, $\mu=5$ and $N=1000$, and the numerical results have been averaged over $1000$ realizations.}
\label{bias_tau}
\end{figure}

The results can be found in Fig \ref{bias_tau}, where we plot the aforementioned observables as a function of $\delta$. In this figure we see that increasing $\delta$ beyond the random value $\delta^\text{rand}$ and using the strategy proposed in Model II to change the biased-dependent topology of the network, results in a significant reduction in the time to reach consensus, while it also results in a significant increase in the probability to reach consensus to the preferred state, when compared to a homogeneous Erd\H{o}s-R\'enyi random network topology of interactions where the establishment of a link is not influenced by the preference of the nodes. Similarly, decreasing $\delta$ below the random value $\delta^\text{rand}$ results in a decrease of the probability to reach consensus on the preferred state and a slight increase in the time to consensus (although the results in this latter case are not conclusive due to the statistical errors). In what concerns the use of the strategy proposed by model I to change the biased-dependent topology of the network, the results indicate that neither the fixation probability $P_1$, nor the consensus time $\tau$ show any significant deviation with respect to the homogeneous Erd\H{o}s-R\'enyi case both for $\delta$ larger or smaller than the random value $\delta^\text{rand}$.

The conclusion we draw from these results, is that what matters the most for the biased group in being more efficient in convincing the rest of the community to reach consensus on their preferred state, is that these agents, on average, have more connections with each other, compared to the random topology of interactions scenario, while at the same time the unbiased group has, on average, less number of connections with each other, compared to the random topology of interactions scenario. To be more specific, from the results concerning the topology variation following the strategy corresponding to model I, we conclude that when both the biased and unbiased agents decide to ``clash", i.e. to interact more among each other at the expense of losing interactions with their peers, then this offers no significant advantage to the biased agents (blue lines in Fig.~\ref{bias_tau} for $\delta>\delta^\text{rand}$). The same conclusion arises when both groups decide to interact less among each other at the expense of winning interactions with their peers (blue lines in Fig.~\ref{bias_tau} for $\delta<\delta^\text{rand}$). On the contrary, when the biased agents decide to interact less with the external group, i.e. the unbiased agents, and more among themselves, and at the same time the unbiased agents interact less among themselves, then at the long run this gives the biased agents an advantage compared to the unbiased agents and results in the preferred opinion being reached faster and more often (green lines in Fig.~\ref{bias_tau} for $\delta>\delta^\text{rand}$). This is because if the group of biased agents is more compact, i.e. when they interact much more among themselves compared to their number of interactions with unbiased agents, and at the same time the unbiased agents are less internally connected, this allows the biased group to have on average a constant bias towards the preferred state for large periods of time, bias that can not be reversed by a weak unbiased group. In this way the biased agents drag the unbiased ones towards their preferred state. 

\section{Conclusions}

In this work, we have studied a variation of the standard classical voter model, where voters have a constant confidence. This means that with a constant probability $p$ they keep their opinion upon an interaction with their neighbor, instead of copying. We assumed that the default confidence is $p=1/2$. However for a fraction $\gamma$ of these voters we assumed that they are biased towards a fixed opinion, in the sense that their confidence when changing from a fixed state, say $s_i=-1$ to $s_i=+1$ is given by $p=(1+v)/2$ with a bias parameter $v$, while the reverse switch $s_i=+1$ to $s_i=-1$ occurs with a confidence equal to $p=(1-v)/2$. We have considered two distinct scenarios in our studies. First, we assumed that there was no dependence of the topology of the network on which the dynamics took place and the type of voters, in which case we studied the model on the complete graph as well as on an Erd\H{o}s-R\'enyi (ER) network. Then we assumed that the topology of interactions of the two distinct type of voters was indeed dependent on their type and we examined strategies that the biased voters could follow to convince faster the rest of the voters to adopt their opinion. 

For the biased-independent topology of interactions, we showed that the fraction of biased agents $\gamma$ and the bias parameter $v$ are not independent parameters, but rather $\gamma v$ is the only relevant parameter. Bias breaks the symmetry of the problem. As a consequence, in the thermodynamic limit the system reaches the absorbing preferred state $m=1,\rho=0$ in a characteristic time $\frac{1}{\gamma v}$. This is at variance with the unbiased voter model ($\gamma v \rightarrow 0$) in which magnetization is conserved and the system remains in a dynamically disordered state with a finite value of $\rho$.
Moving beyond this, we considered finite size effects on the system, for the same biased-independent topology of interactions setup. 
We focused on three observables, the fixation probability $P_1(\sigma)$, or probability to reach the preferred state, as a function of the initial density $\sigma$ of nodes in state $+1$, the average consensus time $\tau(\sigma)$ and the average consensus time to the preferred state $\tau_1(\sigma)$ for which we derive analytical expressions. We show that local effects introduced by the Erd\H{o}s-R\'eny network, as compared with the complete graph case, are accounted by replacing $N$ by $N_{\mu}=\frac{\mu}{\mu+1}N$. The main effect of the bias is to reduce the consensus times so that $\tau$ scales as $\log(N)$ while it scales linearly with $N$ in the limit of no bias ($\gamma v \rightarrow 0$).

Finally, we also have studied the case where the voters lie on Erd\H{o}s-R\'enyi networks of distinct characteristics, i.e. distinct average degrees, depending on the type of the voter. We refer to this as a biased-dependent topology of interactions. In this case, we defined as the parameter that quantifies our deviation from the homogeneous random network, the ratio $\delta=\dfrac{\muBB}{\muUU}$, where $\muXY$ represents the average degree of connections between voters of type X to voters of type Y. With this in mind we identified two candidate rewiring strategies that keep the total average degree $\mu$ constant and we studied how they affect the dynamics to consensus as a function of the parameter $\delta$. In strategy I we considered the case of varying $\muBU$ at the expense of $\muBB$ and $\muUU$. We found that the consensus time and the probability to reach consensus in the preferred option is indifferent to this variation. On the contrary in strategy II we considered the scenario of varying $\muBB$ (and accordingly $\muUU$) while keeping $\muBU$ constant. We found that increasing $\muBB$ resulted in a significant reduction of the consensus time $\tau$, as well as to a significant increment to the probability $P_1$ of reaching consensus to the preferred state. This evidences that what matters the most for the members of the biased group in being more efficient in convincing the rest of the community to reach consensus on their preferred state, is to increase their internal connections within the group and, at the same time, decrease the interactions between the members of the unbiased group. Or, in other words, that a closed community, defined as one where its members have a higher proportion of inside to outside links that it corresponds to a completely random Erd\H{o}s-R\'enyi network, when put into contact with an open one, can lead the rest of the system faster and with higher probability to an agreement to its preferred state. 

{\bfseries Acknowledgments.}
Partial financial support has been received from MCIN/AEI/10.13039/501100011033 and the Fondo Europeo de Desarrollo Regional (FEDER, UE) through project PACSS (RTI2018-093732-B-C21) and the Mar{\'\i}a de Maeztu Program for units of Excellence in R\&D, grant MDM-2017-0711.

\appendix
\setcounter{figure}{0}
\setcounter{equation}{0}
\renewcommand{\thesection}{A\arabic{section}} 		
\renewcommand{\theequation}{A\arabic{equation}} 	
\renewcommand{\thefigure}{A\arabic{figure}} 		
\renewcommand{\thetable}{A\arabic{table}} 

\section*{Appendix}\label{sec:appendix}
To find $T_1(n)$, the solution of the recurrence equation Eq.(\ref{ttrc_d4}) satisfying the boundary conditions $T_{1}(0)=T_{1}(N)=0$ we first define $x(n)$ as 
\begin{equation}
x(n)=T_1(n+1)-T_1(n) \Longleftrightarrow T_1(n)=\sum_{j=0}^{n-1}x(j).
\label{ttrc_dx}
\end{equation}
After substitution in Eq.(\ref{ttrc_d4}), $x(n)$ is found to satisfy the recurrence relation
\begin{equation}
x(n)=ax(n-1)-\frac{1+a}{n}, \,a=\frac{1-\gamma v}{1+\gamma v} \label{ttrc_d8},
\end{equation}
whose solution is
\begin{equation}
x(n)=x(0)a^n-(1+a)L_n(a),\,L_n(a)=\sum_{j=1}^{n}\frac{a^{n-j}}{j},\,L_0(a)=0.
\label{ttrc_d11}
\end{equation}
Replacing in Eq.(\ref{ttrc_dx}) and setting the constant $x(0)$ by imposing the boundary condition $T_1(N)=0$, we get 
\begin{equation}
T_1(n,a)=(1+a)\left[\frac{1-a^n}{1-a^N}\sum_{j=1}^{N-1}L_j(a)-\sum_{j=1}^{n-1}L_j(a)\right].
\label{ttrc_d14}
\end{equation}
The sums of the function $L_j(a)$ can be written as 
\begin{eqnarray}
&\sum_{j=1}^{n-1}L_j(a)=\sum_{j=1}^{n-1}\sum_{q=1}^{j}\frac{a^{j-q}}{q}=\sum_{q=1}^{n-1}\sum_{j=q}^{n-1}\frac{a^{j-q}}{q}\nonumber \\
&=\sum_{q=1}^{n-1}\frac{1}{q}\frac{1-a^{n-q}}{1-a}= \frac{1}{1-a}\left[H_{n-1}-f(n,a)\right]
\label{ttrc_d17}
\end{eqnarray}
where $H_{n}$ (the harmonic function) and $f(n,a)$ are defined by:
\begin{eqnarray}
H_n&=&\sum_{q=1}^n \frac{1}{q},\label{eq:hn}\\
f(n,a)&=&\sum_{q=1}^{n-1}\frac{a^{n-q}}{q},\label{eq:fna}
\end{eqnarray}
which replaced in Eq.(\ref{ttrc_d14}) yields Eq.(\ref{ttrc_d20}) in the main text.

It turns out that the function $f(n,a)$ can be written as 
\begin{equation}
f(n,a)=-\Phi(\frac{1}{a},1,n)-a^n\ln\left(\frac{a-1}{a}\right),
\end{equation}
 in terms of the Lerch transcendent function 
 \begin{equation}\Phi(x,s,n)=\sum_{j=0}^\infty\frac{x^j}{\left(j+n\right)^s}, 
 \end{equation} a relation valid for all values of $a$ and $n$ as the imaginary parts of the logarithm and the Lerch transcendent function cancel out for $a<1$.

\bibliographystyle{unsrt}
\bibliography{BV}

\end{document}